\documentstyle[12pt]{article}
\begin{document}
\author{{\normalsize\bf Yu.A.Markov \thanks{e-mail:markov@icc.ru}
and M.A.Markova}}
\title{Gauge invariance of nonlinear Landau damping rate of Bose excitations
in quark-gluon plasma}
\date{\it Institute of System Dynamics\\
and Control Theory Siberian Branch\\
of Academy of Scienses of Russia,\\
P.O. Box 1233, 664033 Irkutsk, Russia}
\thispagestyle{empty}
\maketitle

\[
{\bf Abstract}
\]

On the basis of the approximate dynamical equations
describing the behavior of
quark-gluon plasma (QGP) in the semiclassical limit
and Yang-Mills equation, the
kinetic equation for longitudinal waves (plasmons)
is obtained. With the Ward identities the gauge
invariance of obtained nonlinear Landau damping rate is proved.
The physical mechanisms defining
nonlinear scattering of a plasmon by QGP particles are
analyzed. The problem on a connection of nonlinear Landau
damping rate of longitudinal oscillations with damping rate,
obtained in the framework of hard thermal loops approximation, is
considered. It is shown that the gauge-dependent part of nonlinear
Landau damping rate for the plasmons with zero momentum vanishes on
mass-shell.
\newpage

{\bf 1. INTRODUCTION}\\

In recent 15-20 years, a theoretical
investigations of properties of quark-gluon plasma has been of great interest.
It is connected with intensive looking for
a QGP in the experiments with collision of ultrarelativistic heavy ions
and application concerning the physics of the early universe.

Two methods to study of the nonequilibrium phenomena in a
QGP are used: method of temperature Green functions
and kinetic approach. Significant progress has been achieved in
the development of the first method. The effective perturbative theory
was constructed in the papers [1] by Pisarski and Braaten, Frenkel and Taylor
on the basis of the resummation of
so-called hard thermal loops (HTL's),
and the problem on the sign and gauge dependence of the
damping rate of the long wavelength excitations in QGP was solved.
Independent solution of "the plasmon puzzle" problem was received by
Kobes, Kunstatter, and Rebhan [2].
The damping rate for heavy quarks interacting with light thermal quarks
and gluons, was found in [3], for soft plasmon and plasmino - in $[4, \,5]$.
The damping rate for energetic fermions and
bosons was derived by Lebedev and Smilga $[6]$ with
extension of the program of resummation outlined in Ref. [1],
by means of inclusion of higher order effects in a hard propagator.
The alternative derivation of damping rate for energetic fermions,
based on introduction of an infrared cutoff was deduced by
Burgess, Marini, and Rebhan [7].
The progress of the thermal QCD makes possible a new
look at the existing of kinetic theory of QGP, developed
by Heinz, Winter, Elze, Vasak and Gyullasy,
Mr\'owczy\'nski and others [8],
and it has given impetus to its further development.

In spite of the fact that the language and methods of these aproaches are very
different, there are close similarities
between HTL approach and transport theory. Originally the
kinetic theory was used by Silin to derive HTL in the photon's
self-energy [9]. HTL's in the quark and gluon self-energies
can be computed similarly. Moreover, Kelly, Liu, Lucchesi
and Manuel [10] have shown that the generating functional of HTL's (with
an arbitrary number of soft external bosonic legs) can be derived
from the classical kinetic theory of QGP. This points to the classical
nature of the hard thermal effects. For hard excitations, the damping rate
has been computed by Heiselberg and Pethick [11] from a
Boltzmann - like equation, with the collision terms included.

A further step in development of kinetic theory was made
by Blaizot and Iancu [12]. In contrast to the early papers on transport
theory of QGP [8], these authors use from outset
the ideas developed in thermal QCD in deriving of the kinetic equations.
The equations obtained by
them on the basis of a truncation of the Schwinger-Dyson hierarchy
isolate consistently the dominant terms in
the coupling constant $g$ in
a set of equations, which describe the response of plasma to weak and
slowly varying disturbances, and encompass all HTL's (conserning
recent investigations in this area, see also papers [13-16]). However, here
it should be noted that if the influence of the average fermionic field
is neglected, then the expression for current induced by
soft gauge fields, obtained in [12]
(and the nonlinear equation of motion, connected with it)
fully
coincide with the corresponding
expression obtained in [10] from usual classical kinetic theory on the
basis of consistent expansion of distribution function in powers of the
coupling constant. This somewhat justifies use of the
(semi)classical kinetic equations found in [8], in spite of
the fact that intermediate approximation schemes in which
these equations were derived, mix leading and nonleading
contributions with respect to the powers of $g$ and
so, are not entirely consistent.

Such close interlacing of two methods of investigation of nonequilibrium
phenomena in QGP leads to the question: can we
calculate the damping rate of soft bosonic modes corresponding to the hard thermal
result [4], remaining in the framework of classical (semiclassical)
kinetic theory only? Xiaofei and Jiarong [17] were the first
to put this question. Because of obtained results, they
have given a positive answer.

As was shown by Heinz and Siemens [18],
in linear approximation the Landau damping is absent in QGP.
In fact, the only mechanism, with that one can connect
the damping following from the kinetic theory with one from
HTL approach, is so-called nonlinear Landau damping. It
bounds up with the nonlinear effects of waves interaction and particles
in QGP. The multiple time-scale method which has proved successful in study of
the nonlinear properties of electromagnetic (Abelian) plasma [19],
was used in [17] for determination of this association. By means of this
method the nonlinear shift of the mass-shell of the longitudinal modes
in the temporal gauge has been obtained by Xiaofei and Jiarong.
Its imaginary part defines required nonlinear
Landau damping rate.
Futher, the limiting expression of the derived damping rate for ${\bf k}=0$-mode
was obtained, and
numerical computations for approximate estimate were performed. The value derived
by this means is in close agreement with similar numerical one obtained
by Braaten and Pisarski [4] in the framework of effective perturbative theory.

However, under close examination of above-mentioned paper we found certain
mistakes in computations, which were of both principle and
nonprinciple character. As it was shown in our early paper [20],
the elimination of these inaccuracies finally leads not only to a numerical
modification of the limiting value of nonlinear
Landau damping rate obtained in [17],
but what is more important, it changes the sign of obtained expression.
This points
to some prematurity of the statements in [17] on obtained connection
between nonlinear Landau damping rate and damping rate from the
HTL-approach.

In this paper, that is further development the ideas outlined in
Ref. [20], we consider the above problem, using the approach
based on obtaining of kinetic
equation for waves in quark-gluon plasma
developed by Kadomtsev, Silin, Tsytovich and others [21] in connection
with ordinary plasma.
We have shown that nonlinear Landau damping rate $\gamma^l({\bf k})$
(which even is not of fixed sign for arbitrary value of ${\bf k}$)
for longitudinal waves in QGP defines two various processes:
the effective spectral pumping of energy from short to long waves
(with complete conservation of excitations energy),
and properly nonlinear dissipation of plasma waves energy in the medium.
The main conclusion of this work is that there is a need to compare
the piece of
$\gamma^l({\bf k})$ that corresponds to nonlinear dissipation of
waves energy and
is a positive for any value of a wave vector ${\bf k}$ and, in particular
for ${\bf k}=0$ - mode, with damping rate of soft boson modes from
HTL-approximation.

The outline of the paper is as follows. In Sec. 2
we derive a system of self-consistent equations
in the covariant gauge
for regular (coherent) and random parts of both the distribution functions of  QGP
particles and gauge field. In Sec. 3
the first order approximation of the colour current is considered
and the correlation function of random oscillations is introduced.
In Sec. 4 we discuss the consistency with gauge symmetry of
used approximation scheme.
In sec. 5 the second and the third orders approximation of the colour
current are studied, and the terms leading in the coupling constant are
separated. In Sec. 6 the kinetic equation for longitudinal waves in
QGP is derived.
In Sec. 7 the nonlinear Landau damping rate is rewritten in the term
of HTL-amplitudes. In Sec. 8 by the effective Ward identities,
the gauge invariance of obtained damping rate is proved.
In Sec. 9 the physical mechanisms defining the nonlinear scattering
of waves by plasma particles are considered.
In Sec. 10 association of the nonlinear Landau damping rate with
damping rate, obtained on the basis of hard thermal loops approximation
is discussed.
In Sec. 11 the estimation
of a value of the nonlinear Landau damping rate of plasmons at rest (at vanishing
three-momentum) is made.
In Sec. 12 it is shown that the gauge-dependent piece of nonlinear
Landau damping rate vanishes on mass-shell.
In Conclusion possible ways of further development of
the scrutinized theory are discussed.
\vspace{1cm}\\
{\bf 2. THE INITIAL EQUATIONS.
THE RANDOM-PHASE APPROXIMATION}\\

We use metric $g^{\mu \nu} = diag(1,-1,-1,-1)$ and choose units such
that $c=k_{B}=1$. The gauge field potentials are $N_{c} \times N_{c}$-matrices
in a color space defined by $A_{\mu}=A_{\mu}^{a}t^{a}$ with $N_{c}^{2}-1$
hermitian generators of $SU(N_{c})$ group in the fundamental representation.
The field strength tensor $F_{\mu \nu}=F_{\mu \nu}^{a}t^{a}$ with
$$
F_{\mu \nu}^{a} = \partial_\mu A_{\nu}^{a} - \partial_\nu A_{\mu}^{a}+
gf^{abc}A_{\mu}^{b}A_{\nu}^{c}
\eqno{(2.1)}
$$
obeys the Yang-Mills (YM) equation in a covariant gauge
$$
\partial_\mu F^{\mu \nu}(X) - ig[A_{\mu}(X),F^{\mu \nu}(X)] -
\xi^{-1} \partial^\nu \partial^\mu A_{\mu}(X) = -j^{\nu}(X),
\eqno{(2.2)}
$$
where $ \xi$ is a gauge parameter. $j^{\nu}$ is the colour current
$$
j^{\nu} = gt^{a} \int d^{4}p \, p^{\nu}[
{\rm Sp} \,t^{a}(f_{q} - f_{\bar{q}})+{\rm Tr} \, (T^{a}f_{g})],
\eqno{(2.3)}
$$
where $T^{a}$ are hermitian generators of $SU(N_{c})$ in the
adjoint representation
$((T^{a})^{bc}=-if^{abc}, {\rm Tr}(T^{a}T^{b})=N_{c} \delta^{ab})$.
We denote the trace over color indices in adjoint
representation as ${\rm Tr}$. Distribution functions of quarks $f_{q}$,
antiquarks $f_{\bar{q}}$, and gluons $f_{g}$ satisfy the dynamical
equations which in the semiclassical limit (neglecting spin effects) are [8]
$$
p^{\mu}{\cal D}_{\mu}f_{q,{\bar{q}}} \pm \frac{1}{2}gp^{\mu} \{ F_{\mu
\nu}, \frac{\partial f_{q,{\bar{q}}}}{\partial p_{\nu}} \} = 0,
\eqno{(2.4)}
$$
$$
p^{\mu} \tilde{\cal D}_{\mu}f_{g}
+ \frac{1}{2}gp^{\mu} \{ {\cal F}_{\mu
\nu}, \frac{\partial f_{g}}{\partial p_{\nu}} \} = 0,
$$
where ${\cal D}_{\mu}$ and $\tilde{\cal D}_{\mu}$ are covariant
derivatives which act as
$$
{\cal D}_{\mu} = \partial_{\mu} - ig[A_{\mu}(X), \cdot \, ],
$$
$$
 \tilde{\cal D}_{\mu} = \partial_{\mu} - ig[{\cal A}_{\mu}(X), \cdot \, ],
$$
[ , ] denotes commutator, $ \{ , \} $ denotes anticommutator, and
${\cal A}_{\mu}$, ${\cal F}_{\mu \nu}$ are defined as ${\cal A}_{\mu}=
A_{\mu}^{a}T^{a}, {\cal F}_{\mu \nu} = F_{\mu \nu}^{a}T^{a}$. Upper sign
in the first equation (2.4) refers to quarks and lower one - to
antiquarks.

We begin with consideration of dynamical equations (2.4).
The distribution functions $f_{q,{\bar{q}}}$ and $f_{g}$
can be decomposed into two parts: regular and random ones, where latter are
generated by spontaneous fluctuations in the plasma
$$
f_{s} = f_{s}^{R} + f_{s}^{T} \; , \; s=q, \bar{q}, g,
\eqno{(2.5)}
$$
so that
$$
\langle f_{s} \rangle = f_{s}^{R} \;, \; \langle f_{s}^{T} \rangle = 0.
\eqno{(2.6)}
$$
Here, angular brackets $ \langle \cdot \rangle$ indicate a statistical
ensemble
of averaging. The initial values of parameters
which characterize the collective degree of a plasma freedom is
such statistical ensemble. For
almost linear collective motion to be considered below it may be
initial values of oscillation phases.

Further we set
$$
A_{\mu} = A_{\mu}^{R} + A_{\mu}^{T} \;,\; \langle A_{\mu}^{T} \rangle = 0,
\eqno{(2.7)}
$$
by definition.
The regular (background) part of the field $A_{\mu}^{R}$
will be considered equal to zero.
The condition for which the last assumption holds, will be closer considered in
Sec. 4.\\

Averaging the equation (2.4) over statistical ensemble, in view of
(2.5)-(2.7), we obtain the equations for the regular parts of the
distribution functions $f_{q, \bar{q}}^{R}$ and $f_{g}^{R}$
$$
p^{\mu} \partial_{\mu} f_{q, \bar{q}}^{R} = igp^{\mu}
\langle [A_{\mu}^{T},f_{q,{\bar{q}}}^{T}] \rangle \mp \frac{1}{2}gp^{\mu}
\langle \{ (F_{\mu \nu}^{T})_{L}, \frac{\partial f_{q, \bar{q}}^{T}}
{\partial p_{\nu}} \} \rangle \mp
\frac{1}{2}gp^{\mu}
\{ \langle (F_{\mu \nu}^{T})_{NL} \rangle ,
\frac{\partial f_{q, \bar{q}}^{R}}
{\partial p_{\nu}} \} \mp
$$
$$
\mp \frac{1}{2}gp^{\mu}
\langle \{ (F_{\mu \nu}^{T})_{NL}, \frac{\partial f_{q, \bar{q}}^{T}}
{\partial p_{\nu}} \} \rangle ,
\eqno{(2.8)}
$$
$$
p^{\mu} \partial_{\mu} f_{g}^{R} = igp^{\mu}
\langle [{\cal A}_{\mu}^{T},f_{g}^{T}] \rangle - \frac{1}{2}gp^{\mu}
\langle \{ ({\cal F}_{\mu \nu}^{T})_{L}, \frac{\partial f_{g}^{T}}
{\partial p_{\nu}} \} \rangle
- \frac{1}{2}gp^{\mu}
\{ \langle ({\cal F}_{\mu \nu}^{T})_{NL} \rangle, \frac{\partial f_{g}^{R}}
{\partial p_{\nu}} \} -
$$
$$
- \frac{1}{2}gp^{\mu}
\langle \{ ({\cal F}_{\mu \nu}^{T})_{NL}, \frac{\partial f_{g}^{T}}
{\partial p_{\nu}} \} \rangle .
$$
Here, indices $"L"$ and $"NL"$ denote the linear and nonlinear parts
with respect to field $A_{\mu}^{a}$ of the strength tensor (2.1).

Subtracting (2.8) from (2.4), we define the equations for
$f_{q, \bar{q}}^{T}$ and $f_{g}^{T}$
$$
p^{\mu} \partial_{\mu} f_{q, \bar{q}}^{T} = igp^{\mu}(
[A_{\mu}^{T},f_{q, \bar{q}}^{T}] -
\langle [A_{\mu}^{T},f_{q, \bar{q}}^{T}] \rangle)
\mp \frac{1}{2}gp^{\mu} \{ (F_{\mu \nu}^{T})_{L},
\frac{\partial f_{q, \bar{q}}^{R}}
{\partial p_{\nu}} \} \mp
$$
$$
\mp  \frac{1}{2}gp^{\mu}
( \{ (F_{\mu \nu}^{T})_{L}, \frac{\partial f_{q, \bar{q}}^{T}}
{\partial p_{\nu}} \} -
\langle \{ (F_{\mu \nu}^{T})_{L}, \frac{\partial f_{q, \bar{q}}^{T}}
{\partial p_{\nu}} \} \rangle) \mp
\frac{1}{2} g p^{\mu}
\{ (F_{\mu \nu}^{T})_{NL} -
\langle (F_{\mu \nu}^{T})_{NL} \rangle , \frac{\partial
f_{q, \bar{q}}^{R}}
{\partial p_{\nu}} \} \mp
$$
$$
\mp \frac{1}{2}gp^{\mu}
( \{ (F_{\mu \nu}^{T})_{NL}, \frac{\partial
f_{q, \bar{q}}^{T}}
{\partial p_{\nu}} \} -
\langle \{ (F_{\mu \nu}^{T})_{NL}, \frac{\partial
f_{q, \bar{q}}^{T}}
{\partial p_{\nu}} \} \rangle ),
\eqno{(2.9)}
$$
$$
p^{\mu} \partial_{\mu} f_{g}^{T} = igp^{\mu}(
[{\cal A}_{\mu}^{T},f_{g}^{T}] -
\langle [{\cal A}_{\mu}^{T},f_{g}^{T}] \rangle)
- \frac{1}{2}gp^{\mu} \{ ({\cal F}_{\mu \nu}^{T})_{L},
\frac{\partial f_{g}^{R}}
{\partial p_{\nu}} \} -
$$
$$
- \frac{1}{2} gp^{\mu}
( \{ ({\cal F}_{\mu \nu}^{T})_{L}, \frac{\partial f_{g}^{T}}
{ \partial p_{\nu}} \} -
\langle \{ ({\cal F}_{\mu \nu}^{T})_{L}, \frac{\partial f_{g}^{T}}
{\partial p_{\nu}} \} \rangle) -
\frac{1}{2} g p^{\mu}
\{ ({\cal F}_{\mu \nu}^{T})_{NL} -
\langle ({\cal F}_{\mu \nu}^{T})_{NL} \rangle , \frac{\partial
f_{g}^{R}}
{\partial p_{\nu}} \} -
$$
$$
- \frac{1}{2}gp^{\mu}
( \{ ({\cal F}_{\mu \nu}^{T})_{NL}, \frac{\partial
f_{g}^{T}}
{\partial p_{\nu}} \} -
\langle \{ ({\cal F}_{\mu \nu}^{T})_{NL}, \frac{\partial
f_{g}^{T}}
{\partial p_{\nu}} \} \rangle ).
$$
The system of equations (2.8) and (2.9) is suitable for
investigation of nonequilibrium processes in QGP such that the
excitation energy of waves is small quantity in relation to the total
energy of particles. In this case it can be used expansion
in powers of oscillations amplitude of the random functions
$f_{s}^{T}$
$$
f_{s}^{T}= \sum_{n=1}^{\infty} f_{s}^{T(n)} \;, \; s=q, \bar{q}, g,
\eqno{(2.10)}
$$
where $f_{s}^{T(n)}$ collects the contributions of
the $n$-th power in $A_{\mu}^{T}$. Substituting expansion (2.10)
into (2.9), and collecting terms of the same order in
$A_{\mu}^{T}$, we derive the system of equations
$$
p^{\mu} \partial_{\mu} f_{q, \bar{q}}^{T(1)} = \mp \frac{1}{2}gp^{\mu}
\{ (F_{\mu \nu}^{T})_{L},
\frac{\partial f_{q, \bar{q}}^{R}}
{\partial p_{\nu}} \},
\eqno{(2.11)}
$$
$$
p^{\mu} \partial_{\mu} f_{q, \bar{q}}^{T(2)} = igp^{\mu}(
[A_{\mu}^{T},f_{q, \bar{q}}^{T(1)}] -
\langle [A_{\mu}^{T},f_{q, \bar{q}}^{T(1)}] \rangle) \mp
\eqno{(2.12)}
$$
$$
\mp \frac{1}{2}gp^{\mu}
( \{ (F_{\mu \nu}^{T})_{L}, \frac{\partial f_{q, \bar{q}}^{T(1)}}
{\partial p_{\nu}} \} -
\langle \{ (F_{\mu \nu}^{T})_{L}, \frac{\partial f_{q, \bar{q}}^{T(1)}}
{\partial p_{\nu}} \} \rangle) \mp
\frac{1}{2} g p^{\mu}
\{ (F_{\mu \nu}^{T})_{NL} -
\langle (F_{\mu \nu}^{T})_{NL} \rangle , \frac{\partial
f_{q, \bar{q}}^{R}}
{\partial p_{\nu}} \} ,
$$
$$
p^{\mu} \partial_{\mu} f_{q, \bar{q}}^{T(3)} = igp^{\mu}(
[A_{\mu}^{T},f_{q, \bar{q}}^{T(2)}] -
\langle [A_{\mu}^{T},f_{q, \bar{q}}^{T(2)}] \rangle) \mp
\frac{1}{2}gp^{\mu}
( \{ (F_{\mu \nu}^{T})_{L}, \frac{\partial f_{q, \bar{q}}^{T(2)}}
{\partial p_{\nu}} \} -
\eqno{(2.13)}
$$
$$
- \langle \{ (F_{\mu \nu}^{T})_{L}, \frac{\partial f_{q, \bar{q}}^{T(2)}}
{\partial p_{\nu}} \} \rangle) \mp
\frac{1}{2}gp^{\mu}
( \{ (F_{\mu \nu}^{T})_{NL}, \frac{\partial
f_{q, \bar{q}}^{T(1)}}
{\partial p_{\nu}} \} -
\langle \{ (F_{\mu \nu}^{T})_{NL}, \frac{\partial
f_{q, \bar{q}}^{T(1)}}
{\partial p_{\nu}} \} \rangle ) \; etc..
$$
Similar equations are obtained for $f_{g}^{T(n)} \;, \; n=1,2,3, \ldots.$

The nonlinear colour current is expressed as
$$
j_{\mu}=j_{\mu}^{R} + j_{\mu}^{T} \;, \; \langle j_{\mu} \rangle =
j_{\mu}^{R} \;, \; j_{\mu}^{T}= \sum_{n=1}^{\infty} j_{\mu}^{T(n)},
\eqno{(2.14)}
$$
where
$$
j_{\mu}^{T(n)}= gt^{a} \int d^{4}p \, p_{\mu}[{\rm Sp} \, t^{a}(f_{q}^{T(n)}-
f_{\bar{q}}^{T(n)}) + {\rm Tr} \, (T^{a}f_{g}^{T(n)})].
\eqno{(2.15)}
$$

Now we turn to the Yang-Mills equation (2.2), connecting the gauge
field with the colour current. Averaging Eq. (2.2) and subtracting
the averaged equation from (2.2) in view of Eqs. (2.7) and (2.14), we
find (for $A_{\mu}^{R}=0$)
$$
\partial_{\mu}(F^{T \mu \nu})_{L} - \xi^{-1} \partial^{\nu}
\partial^{\mu}A_{\mu}^{T} + j^{T(1) \nu} =
\eqno{(2.16)}
$$
$$
= -(j_{NL}^{T \nu} - \langle j_{NL}^{T \nu} \rangle) +
ig \partial_{\mu}([A^{T \mu},A^{T \nu}] - \langle [A^{T \mu},
A^{T \nu}] \rangle )+
$$
$$
+ ig([A^{T}_{\mu},(F^{T \mu \nu})_{L}] - \langle [A_{\mu}^{T},
(F^{T \mu \nu})_{L}] \rangle ) +
g^{2}([A_{\mu}^{T},[A^{T \mu},A^{T \nu}]] -
\langle [A_{\mu}^{T},[A^{T \mu},A^{T \nu}]] \rangle ) .
$$
Here, in the l.h.s. we collect all linear terms with
respect to $A_{\mu}^{T}$ and we denote: $j_{NL}^{T \nu} \equiv
j^{T(2) \nu} + j^{T(3) \nu} + \ldots.$ To account for
nonlinear interaction between waves
and particles in QGP (in first non-vanishing approximation
over the energy of waves), it is sufficiently
to restrict the consideration to third order in powers of $A_{\mu}^T$
in expansion (2.10).

We introduce the following assumption. Eqs. (2.8) represent
the kinetic equations for averaged distribution functions.
The correlation functions in the r.h.s. of these equations
have meaning of the collision terms due to particle-wave interaction
and describe the influence of plasma waves to a background state.
Recently research of similar equations has attracted detailed
attention [16], since on scale large wavelengths of collective
excitations $(\lambda \sim 1/g^2 T)$ they lead to $\mbox{B\"odeker's}$
effective theory [13].
We suppose that a characteristic time of nonlinear relaxation of
the oscillations is small quantity as compared with a time of relaxation
of the distribution particles $f_{s}^{R}$.
Therefore we neglect by change of regular part of the distribution
functions with space and time, assuming that these functions are specified and
describe the global equilibrium in QGP
$$
f_{q, \bar{q}}^{R} \equiv f_{q, \bar{q}}^{0} = 2
\frac{2N_{f} \theta(p_{0})}{(2 \pi)^{3}} \delta(p^{2})
\frac{1}{{\rm e}^{(pu)/T \mp \mu} + 1}, \,
f_{g}^{R} \equiv f_{g}^{0} = 2
\frac{2 \theta(p_{0})}{(2 \pi)^{3}} \delta(p^{2})
\frac{1}{{\rm e}^{(pu)/T } - 1},
\eqno{(2.17)}
$$
where $N_{f}$ - being the number of flavours for massless quarks,
$u_{\mu}$ is the four-velocity of the plasma at temperature $T$,
and $ \mu $ is the quark chemical potential.
\vspace{1cm}\\
{\bf 3. THE LINEAR APPROXIMATION. THE CORRELATION FUNCTION
OF THE RANDOM OSCILLATIONS}\\

We will now come to the derivation of kinetic equation for
waves. The initial equation is Eq. (2.16). The l.h.s.
of Eq. (2.16) contains a linear approximation of the colour current,
explicit form of which is easily defined from Eq. (2.11).
We prefere to work in momentum space; the corresponding equations are
obtained by using
$$
A^{T}_{\mu}(x)= \int d^{4}k A^{T}_{\mu}(k) \, {\rm e}^{-i kX},
$$
and similar translations for $f^{T}_{q, \bar{q}}, \, f^{T}_{g}$. The result of
Fourier transformation for Eq. (2.11) is
$$
f_{q, \bar{q}}^{T(1)}(k,p)= \mp g \frac{\chi^{\nu \lambda}(k,p)}
{pk + ip_{0} \epsilon} \frac{\partial f_{q, \bar{q}}^{0}}
{\partial p^{\lambda}} A^{T}_{\nu}(k) \, , \,
f_{g}^{T(1)}(k,p)= - g \frac{\chi^{\nu \lambda}(k,p)}
{pk + ip_{0} \epsilon} \frac{\partial f_{g}^{0}}
{\partial p^{\lambda}}{\cal A}^{T}_{\nu}(k),
\eqno{(3.1)}
$$
$$
\epsilon \rightarrow +0.
$$
Here $ \chi^{\nu \lambda}(k,p)=(pk)g^{\nu \lambda} - p^{\nu}k^{\lambda}$.
Substituting (3.1) into (2.15) (more precisely, in Fourier transformation
of (2.15), we define a well-known form [8, \,22]
current approximation which is linear with respect to a gauge field
$$
j^{T(1) \mu}(k)= \Pi^{\mu \nu}(k)A^{T}_{\nu}(k),
\eqno{(3.2)}
$$
where
$$
\Pi^{\mu \nu}(k)=g^{2} \int \, d^{4}p \, \frac{p^{\mu}(p^{\nu}(k
\partial_{p})-(kp) \partial^{\nu}_{p}){\cal N}_{eq}}{pk + i p_{0}
\epsilon}
$$
is the high temperature polarization tensor, and
${\cal N}_{eq}= \frac{1}{2} (f_{q}^{0} + f_{\bar{q}}^{0}) + N_{c}
f_{g}^{0}$.

Further we rewrite Eq. (2.16) in the momentum space. Taking into account
(3.2), we obtain
$$
[k^{2}g^{\mu \nu} - (1+ \xi^{-1})k^{\mu}k^{\nu}- \Pi^{\mu \nu}(k)]
A^{Tb}_{\nu}(k)= j_{NL}^{Tb \mu}(k) - \langle j_{NL}^{Tb \mu}(k) \rangle +
$$
$$
+ f^{bcd} \int S^{(I) \mu \nu \lambda}_{k,k_{1},k_{2}}
(A^{Tc}_{\nu}(k_{1})A_{\lambda}^{Td}(k_{2})-
\langle A_{\nu}^{Tc}(k_{1})A_{\lambda}^{Td}(k_{2}) \rangle)
\delta(k-k_{1}-k_{2}) dk_{1}dk_{2}+
\eqno{(3.3)}
$$
$$
+ f^{bcf}f^{fde} \int \Sigma^{(I) \mu \nu \lambda \sigma}
_{k,k_{1},k_{2},k_{3}}
(A^{Tc}_{\nu}(k_{1})A_{\lambda}^{Td}(k_{2})A_{\sigma}^{Te}(k_{3})-
\langle A_{\nu}^{Tc}(k_{1})A_{\lambda}^{Td}(k_{2})A_{\sigma}^{Te}(k_{3})
\rangle)
$$
$$
\delta(k-k_{1}-k_{2}-k_{3}) \, dk_{1}dk_{2}dk_{3},
$$
where
$$
S^{(I) \mu \nu \lambda}_{k,k_{1},k_{2}}=- ig(k^{\nu}
g^{\mu \lambda}+ k_{2}^{\nu}g^{\mu \lambda}- k_{2}^{\mu}g^{\nu \lambda}) \;,
\; \Sigma_{k,k_{1},k_{2},k_{3}}^{(I) \mu \nu \lambda \sigma}=
g^{2}g^{\nu \lambda}g^{\mu \sigma}.
\eqno{(3.4)}
$$

Let us multiply Eq. (3.3) by the complex conjugate
amplitude $A_{\mu}^{T \ast a}(k^{\prime})$ and average it
$$
[k^{2}g^{\mu \nu} - (1+ \xi^{-1})k^{\mu}k^{\nu}- \Pi^{\mu \nu}(k)]
\langle A_{\mu}^{T \ast a}(k^{\prime})A^{Tb}_{\nu}(k) \rangle
= \langle A_{\mu}^{T \ast a}(k^{\prime})j_{NL}^{Tb \mu}(k) \rangle +
$$
$$
+ f^{bcd} \int S^{(I) \mu \nu \lambda}_{k,k_{1},k_{2}}
\langle A_{\mu}^{T \ast a}(k^{\prime})A_{\nu}^{Tc}(k_{1})
A_{\lambda}^{Td}(k_{2}) \rangle
\delta(k-k_{1}-k_{2}) dk_{1}dk_{2}+
\eqno{(3.5)}
$$
$$
+ f^{bcf}f^{fde} \int \Sigma^{(I) \mu \nu \lambda \sigma}
_{k,k_{1},k_{2},k_{3}}
\langle A_{\mu}^{T \ast a}(k^{\prime})A_{\nu}^{Tc}(k_{1})A_{\lambda}^{Td}(k_{2})
A_{\sigma}^{Te}(k_{3}) \rangle
\delta(k-k_{1}-k_{2}-k_{3}) dk_{1}dk_{2}dk_{3}.
$$

We introduce the correlation function of the random oscillations
$$
I_{\mu \nu}^{ab}(k^{\prime},k)= \langle A_{\mu}^{T \ast a}(k^{\prime})
A_{\nu}^{Tb}(k) \rangle.
\eqno{(3.6)}
$$
In conditions of the stationary and homogeneous of QGP,
when the correlation function (3.6) in the coordinate representation
depends on the difference of coordinates and time $ \triangle X=X^{\prime}
- X$ only, we have
$$
I_{\mu \nu}^{ab}(k^{\prime},k)=I_{\mu \nu}^{ab}(k^{\prime})
\delta (k^{\prime}-k).
\eqno{(3.7)}
$$

By the effects of the nonlinear interaction of waves and particles,
the state of QGP becomes weakly inhomogeneous and weakly
nonstationary. The medium nongomogeneity and nonstationary lead to a delta-function
broadering, and $I_{\mu \nu}^{ab}$ depends on both arguments.

Let us introduce $I_{\mu \nu}^{ab}(k^{\prime},k)=I_{\mu \nu}^{ab}
(k, \triangle k)$, $\triangle k = k^{\prime} - k$ with
$\mid\triangle k / k\mid \ll 1$
and insert the correlation function in the Wigner form
$$
I_{\mu \nu}^{ab}(k,x)= \int I_{\mu \nu}^{ab}(k, \triangle k)
{\rm e}^{- i \triangle kx} d \triangle k,
$$
slowly depending on $x$.
In Eq. (3.5) we change $k \rightleftharpoons
k^{\prime} \;, \; a \rightleftharpoons b $, complex conjugate
and subtract obtained equation from Eq. (3.5), beforehand
expanding of the polarization tenzor into Hermitian and anti-Hermitian
parts
$$
\Pi^{\nu \sigma}(k)= \Pi^{H \nu \sigma}(k) + \Pi^{A \nu \sigma}(k) \;,
\; \Pi^{H \nu \sigma}(k)= \Pi^{\ast H \sigma \nu}(k) \;, \;
\Pi^{A \nu \sigma}(k)= - \Pi^{\ast A \sigma \nu}(k).
$$

We assume that anti-Hermitian part of $ \Pi^{A}$ is small in comparison
with $ \Pi^{H}$ and it is a value of the same smallness order, as the
nonlinear terms in the r.h.s.. Therefore it can be suggested
that $ \Pi^{A \nu \sigma}(k) \simeq \Pi^{A \nu \sigma}(k^{\prime})$,
and the term with $ \Pi^{A}$ can be rearranged to the r.h.s..
The remaining terms in the l.h.s.
we expanded in a series in powers of $ \triangle k$ to first smallness order.
This corresponds to ${\it gradient \, expansion}$ procedure usually used
in derivation of kinetic equations $[8, \,12, \,14]$.
Multiplying obtained equation by ${\rm e}^{- i \triangle kx}$
and integrating over $ \triangle k$ with regard to
$$
\int \triangle k_{\lambda} \, I_{\mu \nu}^{ab}(k, \triangle k)
\, {\rm e}^{- i \triangle kx} d \triangle k=
i \frac{\partial I_{\mu \nu}^{ab}(k,x)}{\partial x^{\lambda}},
$$
we obtain finally
$$
\frac{\partial}{\partial k_{\lambda}}[k^{2}g^{\mu \nu}
-(1+ \xi^{-1})k^{\mu}k^{\nu} -
\Pi^{H \mu \nu}(k)] \frac{\partial I_{\mu \nu}^{ab}}{\partial x^{\lambda}}
=2 i \Pi^{A \mu \nu}I_{\mu \nu}^{ab}-
$$
$$
- i \int dk^{\prime} \{
\langle A_{\mu}^{T \ast a}(k^{\prime})j_{NL}^{Tb \mu}(k) \rangle -
\langle A_{\mu}^{Tb}(k)j_{NL}^{\ast Ta \mu}(k^{\prime}) \rangle \} -
$$
$$
- i \{ f^{bcd} \int \, dk^{\prime}dk_{1}dk_{2} \,
S_{k,k_{1},k_{2}}^{(I) \mu \nu \lambda} \langle
A_{\mu}^{T \ast a}(k^{\prime})A_{\nu}^{Tc}(k_{1})A_{\lambda}^{Td}
(k_{2}) \rangle \delta (k-k_{1}-k_{2}) -
$$
$$
- f^{acd} \int \, dk^{\prime}dk_{1}dk_{2} \,
S_{k^{\prime},k_{1},k_{2}}^{\ast (I) \mu \nu \lambda}
\langle
A_{\mu}^{Tb}(k)A_{\nu}^{T \ast c}(k_{1})A_{\lambda}^{T \ast d}
(k_{2}) \rangle \delta (k^{\prime}-k_{1}-k_{2}) \} -
\eqno{(3.8)}
$$
$$
- i \{ f^{bcf}f^{fde} \int \Sigma_{k,k_{1},k_{2},k_{3}}^{(I) \mu \nu \lambda
\sigma} \langle A_{\mu}^{T \ast a}(k^{\prime})A_{\nu}^{Tc}(k_{1})
A_{\lambda}^{Td}(k_{2})A_{\sigma}^{Te}(k_{3}) \rangle
\delta (k-k_{1}-k_{2}-k_{3}) dk_{1}dk_{2}dk_{3}-
$$
$$
-f^{acf}f^{fde} \int \Sigma_{k^{\prime},k_{1},k_{2},k_{3}}
^{\ast (I) \mu \nu \lambda
\sigma} \langle A_{\mu}^{Tb}(k)A_{\nu}^{T \ast c}(k_{1})
A_{\lambda}^{T \ast d}(k_{2})A_{\sigma}^{T \ast e}(k_{3}) \rangle
\delta (k^{\prime}-k_{1}-k_{2}-k_{3}) dk_{1}dk_{2}dk_{3} \},
$$
where $j_{NL}^{Ta \mu}(k)=j^{T(2)a \mu}(k) +
j^{T(3)a \mu}(k)$.

We make several remarks relative to obtained Eq. (3.8). The term
with $\Pi^{A}$ introducing in the r.h.s. of Eq. (3.8) corresponds
to the linear Landau damping. However, as was shown by Heinz and
Siemens [18], linear Landau damping for waves in QGP is absent
and hence this term vanishes.
Futher, the terms with $\Sigma^{(I)}$ can be omitted also.
These terms will be enter into kinetic equation for plasmons (Sec. 5)
with ${\rm Im} \, \Sigma^{(I)}$ that vanishes by reality of the
function $\Sigma^{(I)}$.
\vspace{1cm}\\
{\bf 4. CONSISTENCY WITH GAUGE SYMMETRY}\\

In this Section we shall discuss the consistency of approximation scheme
which we use with requirements of the non-Abelian gauge symmetry.

The initial dynamical equations (2.4) and Yang-Mills equation (2.2)
(without the gauge-fixing condition) transform covariantly under local
transformations
$$
\bar{A}_{\mu} (X) = h (X) ( A_{\mu} (X) + \frac{i}{g} \partial_{\mu} )
h^{\dagger} (X) , \;
h (X) = \exp \, (i \theta^a (X) t^a) ,
$$
with parameter $\theta^a (X)$. We also have transformations of the quark, antiquark
and gluon distribution functions [8]
$$
\bar{f}_{q, \bar{q}} (p, X) = h (X) f_{q, \bar{q}} (p, X) h^{\dagger} (X) , \;
\bar{f}_{g} (p, X) = H (X) f_{g} (p, X) H^{\dagger} (X) ,
$$
where $H^{ab} (X) = {\rm Sp} [t^a h(X) t^b h^{\dagger} (X)]$.

As known (see, e.g. [16]), after the splitting of (2.5), (2.7)
the resulting equations left two symmetries, the ${\it background \; gauge \;
symmetry}$,
$$
\bar{A}_{\mu}^{R} (X) = h (X) ( A_{\mu}^{R} (X) + \frac{i}{g} \partial_{\mu} )
h^{\dagger} (X) , \;
\bar{A}_{\mu}^{T} (X) = h (X) A_{\mu}^{T} (X) h^{\dagger} (X) ,
\eqno{(4.1)}
$$
and the ${\it fluctuation \; gauge \; symmetry}$,
$$
\bar{A}_{\mu}^{R} (X) = 0 , \;
\bar{A}_{\mu}^{T} (X) = h (X) ( A_{\mu}^{R} (X) + \bar{A}_{\mu}^{T} (X) +
\frac{i}{g} \partial_{\mu} )
h^{\dagger} (X) .
\eqno{(4.2)}
$$

The condition which we impose on regular part of a gauge field $A_{\mu}^{R}$ and
requirement that the statistical average of the fluctuation vanishes
$\langle A_{\mu}^{T} \rangle = 0$, break both of types symmetry (4.1)
and (4.2). Thus in the case of a gauge transformation (4.1)
we obtain $\bar{A}_{\mu}^{R} \neq 0$, and in the case of (4.2)
we come to noninvariance of the constraint
$\langle A_{\mu}^{T} \rangle = 0$.
Moreover, introduced correlation function (3.6) also have explicitly
a gauge noncovariant character. This leads to the fact that
calculations in the preceding Sections are gauge
noncovariant, and therefore their value is doubted.

Nevertheless, there is the special case, when preceding (and following)
conclusions are justified. This is the case of a colourless fluctuation,
for which $I^{ab}_{\mu \nu}(k,x)= \delta^{ab}I_{\mu \nu}(k,x)$.
We can obtain a gauge invariant equation for $I_{\mu \nu}(k,x)$ only
in this restriction, in spite of the fact that the intermediate calculations
spoil non-Abelian gauge symmetry of the initial equations
(2.2)-(2.4).

In principle, we shall be  able to maintain explicit
background gauge symmetry (4.1) at each step of our calculations,
as it has been done, for example, by Blaizot and Iancu [14] for derivation
of the Boltzmann equation describing the relaxation of ultrasoft
$( \lambda \sim 1/g^2T)$ colour excitations.
First of all we assume that
$A^{R}_{\mu} \neq 0$. Then as the gauge-fixing condition for the random field
$A^{T}_{\mu}$, we choose the background field gauge
$$
{\cal D}^{R}_{\mu}(X) A^{R \, \mu}(X) =0, \;
{\cal D}^{R}_{\mu}(X) \equiv \partial_{\mu} - igA^{R}_{\mu}(X),
\eqno{(4.3)}
$$
which is manifestly covariant with respect to the gauge transformations
of the background gauge field $A^{R}_{\mu}(X)$. We lastly define a gauge
covariant Wigner function as in Refs. $[8, \,14]$
$$
\acute{I}^{ab}_{\mu \nu}(k,x)=
\int \, \acute{I}^{ab}_{\mu \nu}(s,x) \, {\rm e}^{iks} {\rm d}s, \;
s \equiv X_1 - X_2, \, x \equiv \frac{1}{2}(X_1 + X_2),
$$
where
$$
\acute{I}^{ab}_{\mu \nu}(s,x) \equiv
U^{a a^{\prime}}(x,x + \frac{s}{2}) \,
I^{a^{\prime}b^{\prime}}(x + \frac{s}{2},x - \frac{s}{2}) \,
U^{b^{\prime}b}(x - \frac{s}{2},x),
$$
instead of the usual Wigner function $I_{\mu \nu}^{a b} (k, x)$,
whose "poor" transformation properties follow from initial definition
$I^{ab}_{\mu \nu}(X_1,X_2) = \langle
A^{T a}_{\mu}(X_1)A^{T b}_{\nu}(X_2) \rangle$.
The function $U(x,y)$ is the non-Abelian parallel transporter
$$
U(x,y) = {\rm P}
\exp \Big\{ -ig \int_{\gamma} {\rm d}z^{\mu} A^{R}_{\mu}(z) \Big\}.
$$
The path $\gamma$ is the straight line joining $x$ and $y$.

The derivation of the kinetic equation for plasmons in this approach
becomes quite cumbersome and non-trivial. For example, in the l.h.s. of
equations for random parts of distributions
(2.11)-(2.13), the covariant derivative ${\cal D}^{R}_{\mu}$
will be used instead of the ordinary one $\partial_{\mu}$.
Besides, we cannot suppose that regular parts of distributions functions
are specified and equal to equilibrium Fermi-Dirac and Bose-Einstein
distributions (2.17). It is necessary also take
into account their change using kinetic equations (2.8) with
the collision terms in the r.h.s. of (2.8), which describe the reaction
of the soft fluctuation on the background distributions.
The correlators in the r.h.s. of Eq. (2.8) can be expressed in terms
of the function $I_{\mu \nu}^{a b}$ and the distributions of hard particles
$f_{s}^{R} (p, X) , s = q, \bar{q}, g$ , only.

However, if we restrict our consideration to study of colourless
excitations and replace
the distribution functions of hard particles by their equilibrium
values (2.17), then it leads to effective vanishing of terms
explicitly contained the mean field $A^{R}_{\mu}$.
This follows e.g., from the analysis of derivation of Boltzmann equation
by Blaizot and Iancu [14].
Therefore the most simple way of derivation of kinetic equations for
soft colourless QGP excitations is assume $A^{R}_{\mu} = 0$
and use prime gauge noncovariant correlator (3.6).
In this case the background field gauge (4.3) is reduced to
a covariant one.
The kinetic equation obtained by this means is gauge invariant if all contributions
at the leading order in $g$ to the probability of nonlinear wave scattering
by hard thermal particles are taken into account (see also discussion
in Conclusion).

In fact, the requirement of nontriviality of a colour structure of Wigner
function $I_{\mu \nu}^{a b} (k, x)$ leads to necessity of existence of
nonvanishing mean field $A_{\mu}^{R}$ and/or the external colour
current and conversely, the existence of mean field in QGP
results in colour structure $I_{\mu \nu}^{a b} (k, x)$, which is different
from the identity.
\vspace{1cm}\\
{\bf 5. THE SECOND AND THIRD APPROXIMATIONS OF THE COLOUR CURRENT}\\

Now we conserned with computation of the nonlinear corrections
to the current in the r.h.s. of basic Eq. (3.8).

At first we define $f_{q, \bar{q}}^{T(2)}$. We carry out the Fourier
transformation of Eq. (2.12) and substituting
the obtained $f^{T(1)}_{q, \bar{q}}$ from (3.1) into derived expression,
we find
$$
f_{q, \bar{q}}^{T(2)}= \mp g^{2} \frac{[t^{b},t^{c}] p^{\nu}p^{\lambda}}
{pk + ip_{0} \epsilon} \int \, \frac{(k_{2} \partial_{p} f_{q, \bar{q}}
^{0})}{pk_{2} + ip_{0} \epsilon}
(A_{\nu}^{b}(k_{1})A_{\lambda}^{c}(k_{2})
- \langle A_{\nu}^{b}(k_{1})A_{\lambda}^{c}(k_{2}) \rangle)
\delta (k - k_{1} - k_{2}) dk_{1}dk_{2} +
\eqno{(5.1)}
$$
$$
+ \frac{g^{2}}{2} \frac{ \{ t^{b},t^{c} \} }
{pk + ip_{0} \epsilon} \int \, \chi^{\nu \lambda}(k_{1},p)
\frac{\partial}{\partial p^{\lambda}} \left(
\frac{\chi ^{\sigma  \rho}(k_{2},p)}{pk_{2} + ip_{0} \epsilon} \,
\frac{\partial f_{q, \bar{q}}^{0}}{\partial p^{\rho}} \right)
(A_{\nu}^{b}(k_{1})A_{\sigma}^{c}(k_{2})
- \langle A_{\nu}^{b}(k_{1})A_{\sigma}^{c}(k_{2}) \rangle )
$$
$$
\delta (k - k_{1} - k_{2}) \, dk_{1}dk_{2}.
$$
From here on the suffix "T" for a gauge field is omitted.
The expression for $f_{g}^{T(2)}(k,p)$ is obtained from (5.1)
by choosing upper sign and replacements $f_{q}^{0} \rightarrow f_{g}^{0},
t^{a} \rightarrow T^{a}$. Substituting obtained expressions $f^{T(2)}_{s},
s=q, \bar{q},g$ into (2.15) (for $n=2$), we find required current correction
$$
j^{T(2)a \mu}(k)= -ig^{3}f^{abc} \int \, d^{4}p \,
\frac{p^{\mu}p^{\nu}p^{\lambda}}
{pk + ip_{0} \epsilon} \, \frac{(k_{2} \partial_{p} {\cal N}_{eq}
)}{pk_{2} + ip_{0} \epsilon} \,
(A_{\nu}^{b}(k_{1})A_{\lambda}^{c}(k_{2})
- \langle A_{\nu}^{b}(k_{1})A_{\lambda}^{c}(k_{2}) \rangle)
\eqno{(5.2)}
$$
$$
\delta (k - k_{1} - k_{2}) \, dk_{1}dk_{2} +
$$
$$
+ \frac{g^{3}}{4}d^{abc} \int \, d^{4}p \, \frac{p^{\mu} \chi^{\nu \lambda}
(k_{1},p)}{pk + ip_{0} \epsilon} \frac{\partial}{\partial
p^{\lambda}} \left(
\frac{\chi ^{\sigma  \rho}(k_{2},p)}{pk_{2} + ip_{0} \epsilon} \,
\frac{\partial (f_{q}^{0}-f_{\bar{q}}^{0})}{\partial p^{\rho}} \right)
(A_{\nu}^{b}(k_{1})A_{\sigma}^{c}(k_{2})-
\langle A_{\nu}^{b}(k_{1})A_{\sigma}^{c}(k_{2}) \rangle)
$$
$$
\delta (k - k_{1} - k_{2}) \, dk_{1}dk_{2}.
$$
The contribution of gluons to the expression with symmetric structure
constant $d^{abc}$ here drops out. This is connected with
the fact that in calculation of trace of anti-commutators we have:
${\rm Sp} \, t^{a}
\{ t^{b},t^{c} \}= \frac{1}{2}d^{abc}$ - for quarks and antiquarks, and
${\rm Tr} \, T^{a} \{ T^{b},T^{c} \} =0$ - for gluons. The symmetry of
contributions can be restored if we note that besides usual gluon
current $j_{g}^{\mu}(x)= gt^{a} \int \, d^{4}p \, p^{\mu} {\rm Tr}
(T^{a}f_{g}(x,p))$, the dynamical equation for gluons admits
a covariant conserving quantity
$$
\zeta gt^{a} \int \, d^{4}p \, {\rm Tr}({\cal P}^{a}f_{g}(x,p)),
\eqno{(5.3)}
$$
where $({\cal P}^{a})^{bc}=d^{abc}$ and $ \zeta$ is a certain arbitrary
constant. The covariant continuity of (5.3) is evident from the identity:
$[{\cal P}^{a},T^{b}]=if^{abc} {\cal P}^{c}$. On addition of (5.3) to
(2.3) we have contributions to the nonlinear current corrections only.
Adding (5.3) to the second current iteration (2.15) and taking into
account the equality
$$
{\rm Sp} \, {\cal P}^{a} \{ T^{b},T^{c} \}= N_{c}d^{abc},
$$
we derive more general expression for $j^{T(2)}$,
instead of (5.2)
$$
j^{T(2)a \mu}(k)= \int \, S^{abc \mu \nu \lambda}_{k,k_{1},k_{2}}
(A^{b}_{\nu}(k_{1})A_{\lambda}^{c}(k_{2})-
\langle A^{b}_{\nu}(k_{1})A_{\lambda}^{c}(k_{2}) \rangle)
\delta (k - k_{1} - k_{2}) dk_{1}dk_{2},
\eqno{(5.4)}
$$
where $\, S_{k,k_{1},k_{2}}^{abc \mu \nu \lambda}=
f^{abc}S_{k,k_{1},k_{2}}^{(II) \mu \nu \lambda}+
d^{abc}S_{k,k_{1},k_{2}}^{(III) \mu \nu \lambda}$,
$$
S_{k,k_{1},k_{2}}^{(II) \mu \nu \lambda}=
-ig^{3} \int \, d^{4}p \, \frac{p^{\mu}p^{\nu}p^{\lambda}}
{pk + ip_{0} \epsilon} \, \frac {(k_{2} \partial_{p} {\cal N}_{eq})}
{pk_{2} + ip_{0} \epsilon},
\eqno{(5.5)}
$$
$$
S_{k,k_{1},k_{2}}^{(III) \mu \nu \lambda}=
\frac{g^{3}}{2} \int \, d^{4}p \, \frac{p^{\mu} \chi^{\nu \sigma}(k_{1},p)}
{pk + ip_{0} \epsilon} \, \frac{\partial}{\partial p^{\sigma}}
\left( \frac {\chi^{\lambda \rho}(k_{2},p)}
{pk_{2} + ip_{0} \epsilon} \, \frac{\partial \tilde{{\cal N}}_{eq}}
{\partial p^{\rho}} \right),
\eqno{(5.6)}
$$
$$
\tilde{\cal N}_{eq}= \frac{1}{2}(f_{q}^{0} - f_{\bar{q}}^{0}) +
\zeta N_{c}f_{g}^{0}.
$$

The tensor structure of $S^{(III) \mu \nu \lambda}_{k,k_{1},k_{2}}$
exactly coincides with appropriate expression obtained in
calculation of $j^{T(2) \mu}$ in Abelian plasma [21],
and hence piece of current with $d^{abc}$ has
a meaning of Abelian part of the colour current $j^{T(2)a \mu}$.
The term with $S_{k,k_{1},k_{2}}^{(II) \mu \nu \lambda}$
is purely non-Abelian, i.e. it has no Abelian counterpart.

Let us estimate orders of $S^{(II)}$ and $S^{(III)}$. Following
usual terminology [1], we call an energy or a momentum "soft"
when it is of order $gT$, and "hard" when it is of order $T$. We
will be considered, as in Ref. [12], that collective excitations
carrying soft momenta, i.e. $k \sim gT$, and plasma particles
have the typical hard energies: $p \sim T$. In a coordinate representation
the first of conditions denotes that oscillation amplitude $A^{a}_{\mu}$
and distribution functions of hard particles $f_{s}(X,p), s=q, \bar{q},g$
change on the scale $X \sim 1/gT$.
On this basis, we have the
following estimate for $S^{(II)}$
$$
S_{k,k_{1},k_{2}}^{(II) \mu \nu \lambda} \sim g^{2}T.
\eqno{(5.7)}
$$
Here, we considered that by virtue of the definitions (2.17)
${\cal N}_{eq} \sim 1/T^{2}$.

In expression (5.6) the integral of energy with gluon distribution
function is infrared divergent. In a similar manner [17], we regulate it
by introducing an electric mass cutoff of order $gT$, and only take
the leading term in $g$. Then it can be found that in (5.6), the
part related to the gluon distribution function is of order $g^{2}T
(g \ln g)$ and the other part related to the quark and antiquark
distribution functions is of order $g^{3}T$, i.e.
$$
S_{k,k_{1},k_{2}}^{(III) \mu \nu \lambda} \sim g^{2}T(g \ln g)+
g^{3}T.
$$

Hence, $S^{(II)}$ which is purely non-Abelian, is of lower order in
the coupling constant than $S^{(III)}$, which has an Abelian
counterpart. This fact was first seen in Ref. [17].

The expression for a colour current in third order in field
is defined by means of reasoning similar previous ones. Performing
the Fourier transformation of equation (2.13), taking into account
(3.1), (5.1) and equalities
$$
{\rm Tr} \, ( \{ T^{a},T^{b} \} \{ T^{d},T^{e} \} )=
N_{c}d^{abc}d^{cde} + 4 \delta^{ab} \delta^{ed} + 2 \delta^{ad}
\delta^{eb} + 2 \delta^{ae} \delta^{bd},
\eqno{(5.8)}
$$
$$
{\rm Tr} \, ( \{ {\cal P}^{a},T^{b} \} \{ T^{d},T^{e} \} )=0,
$$
we find the required form of $j^{T(3)a \mu}$
$$
j^{T(3)a \mu}(k)= \int \, \Sigma_{k,k_{1},k_{2},k_{3}}^{abde \mu
\nu \lambda \sigma}(
A_{\nu}^{b}(k_{3})A_{\lambda}^{d}(k_{1})A_{\sigma}^{e}(k_{2})-
A_{\nu}^{b}(k_{3}) \langle A_{\lambda}^{d}(k_{1})
A_{\sigma}^{e}(k_{2}) \rangle -
$$
$$
- \langle A_{\nu}^{b}(k_{3})A_{\lambda}^{d}(k_{1})A_{\sigma}^{e}(k_{2}) \rangle )
\delta (k - k_{1} - k_{2} - k_{3}) \, dk_{1}dk_{2}dk_{3} +
\eqno{(5.9)}
$$
$$
+ d^{abc}f^{cde} \int \, R_{k,k_{1},k_{2},k_{3}}^{ \mu
\nu \lambda \sigma}(
A_{\nu}^{b}(k_{3})A_{\lambda}^{d}(k_{1})A_{\sigma}^{e}(k_{2})-
\langle A_{\nu}^{b}(k_{3})A_{\lambda}^{d}(k_{1})A_{\sigma}^{e}(k_{2}) \rangle )
$$
$$
\delta (k - k_{1} - k_{2} - k_{3}) \, dk_{1}dk_{2}dk_{3}.
$$
Here,
$$
\Sigma_{k,k_{1},k_{2},k_{3}}^{abde \mu \nu \lambda \sigma}=
f^{abc}f^{cde} \Sigma_{k,k_{1},k_{2},k_{3}}^{(II) \mu \nu \lambda
\sigma}+
f^{abc}d^{cde} \Sigma_{k,k_{1},k_{2},k_{3}}^{(III) \mu \nu \lambda
\sigma}+
\delta^{ab} \delta^{de} \Sigma_{k,k_{1},k_{2},k_{3}}^{(IV) \mu \nu \lambda
\sigma}+
d^{abc}f^{cde} \Sigma_{k,k_{1},k_{2},k_{3}}^{(V) \mu \nu \lambda
\sigma} +
$$
$$
+ d^{abc}d^{cde} \Sigma_{k,k_{1},k_{2},k_{3}}^{(VI) \mu \nu \lambda
\sigma}+
( \delta^{ab} \delta^{de} + \delta^{ad} \delta^{be} + \delta^{ae} \delta^{db})
\Sigma_{k,k_{1},k_{2},k_{3}}^{(VII) \mu \nu \lambda \sigma},
$$
$$
\Sigma_{k,k_{1},k_{2},k_{3}}^{(II) \mu \nu \lambda \sigma} =
-g^{4} \int \, d^{4}p \,
\frac{p^{\mu}p^{\nu}p^{\lambda}p^{\sigma}}{pk + ip_{0} \epsilon}
\, \frac{1}{p(k_{1} +k_{2}) + ip_{0} \epsilon}
\, \frac{(k_{2} \partial_p {\cal N}_{eq})}{pk_{2} + ip_{0} \epsilon},
\eqno{(5.10)}
$$
$$
\Sigma_{k,k_{1},k_{2},k_{3}}^{(IV) \mu \nu \lambda \sigma}=
- \frac{g^{4}}{2N_{c}} \int \, d^{4}p \frac{p^{\mu} \chi^{\nu \tau}(k_{3},p)}
{pk + ip_{0} \epsilon}
\frac{\partial}{\partial p^{\tau}} \Big(
\frac{\chi^{\lambda \alpha}(k_{1},p)}{p(k_{1} + k_{2}) + ip_{0} \epsilon}
\, \frac{\partial}{\partial p^{\alpha}} \Big(
\frac{\chi^{\sigma \rho}(k_{2},p)}{pk_{2} + ip_{0} \epsilon}
\frac{\partial {\cal N}_{eq}}{\partial p^{\rho}} \Big) \Big),
\eqno{(5.11)}
$$
$$
\Sigma_{k,k_{1},k_{2},k_{3}}^{(VI) \mu \nu \lambda \sigma}=
\frac{N_{c}}{2} \Sigma_{k,k_{1},k_{2},k_{3}}^{(IV) \mu \nu \lambda \sigma}.
$$
The expression for $ \Sigma^{(VII)}$ is obtained from (5.11) by exception
of quark and antiquark contributions. The availability of the term with
$ \Sigma^{(VII)}$ is reflection of more complicated colour structure
of the gluon kinetic equation in comparison with quark and antiquark equations,
that manifests here, in appearance of additional terms in (5.8) as
compared with
$$
{\rm Sp}( \{ t^a,t^b \} \{t^d,t^e \}) =
\frac{1}{2}d^{abc}d^{cde} + \frac{1}{N_c} \delta^{ab} \delta^{de}.
$$

The terms with $ \Sigma^{(III)}, \Sigma^{(V)}$ and $R$ are defined as
the interference of Abelian and non-Abelian contributions.
For colourless fluctuations of QGP, which we study in a given paper,
the correlation function (3.6) is proportional to the
identity. This leads to the absence of the interference of Abelian and
non-Abelian contributions. For this reason their explicit form is
not given here.

At the end of this Section we estimate the order of $ \Sigma^{(II)}$ and
$ \Sigma^{(IV)}$. It follows from the expression (5.1) that
$$
\Sigma_{k,k_{1},k_{2},k_{3}}^{(II) \mu \nu \lambda \sigma}
\sim g^{2}.
\eqno{(5.12)}
$$
Cutting off, as in the previous Section, integration limit for the gluon
distribution function, we find
$$
\Sigma^{(IV)} \sim \Sigma^{(VI)} \sim g^{3} + g^{4} \;,
\; \Sigma^{(VII)} \sim g^{3}.
$$
By this means, purely non-Abelian contribution of $ \Sigma^{(II)}$ is
of lower order in the coupling constant than Abelian - $ \Sigma^{(IV)},
\Sigma^{(VI)}$ and $ \Sigma^{(VII)}$.
\vspace{1cm}\\
{\bf 6. THE KINETIC EQUATION FOR LONGITUDINAL WAVES}\\

Now we turn to initial equation for waves (3.8). We substitute obtained
nonlinear corrections of induced current by field (5.4) and
(5.9) into this equation.

Because of a nonlinear wave interaction  the phase correlation effects
take plays.
By virtue of their smallness, fourth-order correlators can
be approximately divided into product of the correlation functions
$\langle A^{\ast}(k^{\prime}) A(k) \rangle$.
For third-order correlation functions this decomposition vanishes,
and it should be considered a weak correlation of phases fields.
For this purpose we use the nonlinear equation of a field (3.3),
taking into account only the terms of the second
order in  $A$, in the r.h.s. of (3.3)
$$
[k^{2}g^{\mu \nu} - (1 + \xi^{-1})
k^{\mu}k^{\nu} - \Pi^{\mu \nu}(k)]A_{\nu}^{a}(k)=
$$
$$
= f^{abc} \int \, S_{k,k_{1},k_{2}}^{\mu \nu \lambda}
(A_{\nu}^{b}(k_{1})A_{\lambda}^{c}(k_{2}) -
\langle A_{\nu}^{b}(k_{1})A_{\lambda}^{c}(k_{2}) \rangle)
\delta(k - k_{1} - k_{2}) dk_{1}dk_{2}.
\eqno{(6.1)}
$$
The approximate solution of this equation is in the form
$$
A_{\mu}^{a}(k)=A_{\mu}^{(0)a}(k) - {\cal D}_{\mu \nu}(k)f^{abc} \int \,
S_{k,k_{1},k_{2}}^{\nu \lambda \sigma}(
A_{\lambda}^{(0)b}(k_{1})A_{\sigma}^{(0)c}(k_{2}) -
$$
$$
- \langle A_{\lambda}^{(0)b}(k_{1})A_{\sigma}^{(0)c}(k_{2}) \rangle )
\delta(k - k_{1} - k_{2}) dk_{1}dk_{2},
\eqno{(6.2)}
$$
where $S_{k, k_1, k_2}^{\mu \nu \lambda} \equiv
       S_{k, k_1, k_2}^{(I) \mu \nu \lambda} +
       S_{k, k_1, k_2}^{(II) \mu \nu \lambda}$ and
$A_{\mu}^{(0)a}(k)$ is a solution of homogeneous Eq. (6.1)
corresponding free fields, and
$$
{\cal D}_{\mu \nu}(k)=-[k^{2}g_{\mu \nu} - (1 + \xi^{-1})k_{\mu}k_{\nu} -
\Pi_{\mu \nu}(k)]^{-1}
\eqno{(6.3)}
$$
represents the medium modified (retarded) gluon propagator.

Now we substitute (6.2) into third-order correlation functions
entering to Eq. (3.8). Because $A^{(0)}$ represents amplitudes of entirely
uncorrelated waves, the correlation function with three $A^{(0)}$ drops out.
In this case
every term in $ \langle A_{\mu}^{\ast a}(k^{\prime})A_{\nu}^{c}
(k_{1})A_{\lambda}^{d}(k_{2}) \rangle $ and
$ \langle A_{\mu}^{b}(k)A_{\nu}^{\ast c}
(k_{1})A_{\lambda}^{\ast d}(k_{2}) \rangle $ should be defined more
exactly. In the correlation
functions of four amplitudes, within the accepted accuracy, it can
not make distinctions between the fields $A$ and $A^{(0)}$.

Finally Eq. (3.8) becomes
$$
\frac{\partial}{\partial k_{\lambda}}[k^{2}g^{\mu \nu} -
(1 + \xi ^{-1})k^{\mu}k^{\nu} - \Pi^{H \mu \nu}(k)]
\frac{\partial I_{\mu \nu}^{ab}}{\partial x^{\lambda}}=
$$
$$
= -i \int \, dk^{\prime} dk_{1} dk_{2} dk_{3} \{
f^{bcf}f^{fde} \delta(k - k_{1} - k_{2} - k_{3}) \tilde{\Sigma}_{k,
k_{1},k_{2},k_{3}}^{\mu \nu \lambda \sigma}( \langle
A_{\mu}^{\ast a}(k^{\prime})A_{\nu}^{c}(k_{3})A_{\lambda}^{d}(k_{1})
A_{\sigma}^{e}(k_{2}) \rangle -
$$
$$
- \langle A_{\mu}^{\ast a}(k^{\prime})A_{\nu}^{c}(k_{3}) \rangle
\langle A_{\lambda}^{d}(k_{1})A_{\sigma}^{e}(k_{2}) \rangle ) -
$$
$$
- f^{acf}f^{fde} \delta(k^{\prime} - k_{1} - k_{2} - k_{3})
\tilde{\Sigma}_{k^{\prime},
k_{1},k_{2},k_{3}}^{\ast \mu \nu \lambda \sigma}( \langle
A_{\mu}^{b}(k)A_{\nu}^{\ast c}(k_{3})A_{\lambda}^{\ast d}(k_{1})
A_{\sigma}^{\ast e}(k_{2}) \rangle -
$$
$$
- \langle A_{\mu}^{b}(k)A_{\nu}^{\ast c}(k_{3}) \rangle
\langle A_{\lambda}^{\ast d}(k_{1})A_{\sigma}^{\ast e}(k_{2}) \rangle ) \} +
\eqno{(6.4)}
$$
$$
+ if^{bcd}f^{aef} \int \, dk^{\prime} \int \, dk_{1}dk_{2}
dk^{\prime}_{1}dk^{\prime}_{2} \, ({\cal D}^{\ast}_{\rho \alpha}(k^{\prime}) -
{\cal D}_{\alpha \rho}(k)) S_{k,k_{1},k_{2}}^{\rho \mu \nu}
S_{k^{\prime},k_{1}^{\prime},k_{2}^{\prime}}^{\ast \alpha \lambda \sigma} (
\langle A_{\mu}^{c}(k_{1})A_{\nu}^{d}(k_{2})
$$
$$
A_{\lambda}^{\ast e}(k_{1}^{\prime})A_{\sigma}^{\ast f}(k_{2}^{\prime})
\rangle - \langle A_{\mu}^{c}(k_{1})A_{\lambda}^{d}(k_{2}) \rangle
\langle A_{\lambda}^{\ast e}(k_{1}^{\prime})
A_{\sigma}^{\ast f}(k_{2}^{\prime}) \rangle )
\delta (k - k_{1} - k_{2}) \delta (k^{\prime} - k_{1}^{\prime}
- k_{2}^{\prime}).
$$
Here, we keep the terms of leading order in $g$ only and set
$$
\tilde{\Sigma}_{k,k_{1},k_{2},k_{3}}^{\mu \nu \lambda \sigma}
\equiv \Sigma_{k,k_{1},k_{2},k_{3}}^{(II) \mu \nu \lambda \sigma} -
(S_{k,k_{3},k_{1} + k_{2}}^{\mu \nu \rho} -
S_{k,k_{1} + k_{2},k_{3}}^{\mu \rho \nu}){\cal D}_{\rho \alpha}(k_{1}
+k_{2})S_{k_{1} + k_{2},k_{1},k_{2}}^{\alpha \lambda \sigma}.
\eqno{(6.5)}
$$

It follows from the definition (6.3) that the propagator is of the order
$\sim 1/g^{2}T^{2}$. Taking into account (3.4), (5.7)
and (5.12), we see that
all terms in the r.h.s. of (6.4) are of the same order. This
explains the fact that in the expansion of the current (2.14) the following term
- $j^{T(3)}$ should be retained in addition to the first nonlinear
correction $j^{T(2)}$: it leads to the effects of the same order of magnitude.

Let us make the correlation decoupling of the fourth-order correlators
in the r.h.s. of Eq. (6.4)
in the terms of the pair ones by the rule
$$
\langle A(k_{1})A(k_{2})A(k_{3})A(k_{4}) \rangle =
\langle A(k_{1})A(k_{2}) \rangle \langle A(k_{3})A(k_{4}) \rangle +
\langle A(k_{1})A(k_{3}) \rangle \langle A(k_{2})A(k_{4}) \rangle +
$$
$$
+ \langle A(k_{1})A(k_{4}) \rangle \langle A(k_{2})A(k_{3}) \rangle.
$$

Taking into account that the spectral densities in the r.h.s.
of Eq. (6.4) can be considered as stationary and homogeneous those,
i.e. having the form (3.7), and setting $I_{\mu \nu}^{ab}= \delta^{ab} I_{\mu \nu}$, we find instead of
Eq. (6.4)
$$
\frac{\partial}{\partial k_{\lambda}}[k^{2}g^{\mu \nu} - (1 + \xi^{-1})
k^{\mu}k^{\nu} - \Pi^{H \mu \nu}(k)]
\frac{\partial I_{\mu \nu}}{\partial x^{\lambda}} =
$$
$$
= 2N_{c} \int \, dk_{1} \, {\rm Im}(
\tilde{\Sigma}_{k,k,k_{1},- k_{1}}^{\mu \nu \lambda \sigma} -
\tilde{\Sigma}_{k,k_{1},k,- k_{1}}^{\mu \nu \sigma \lambda})
I_{\mu \lambda}(k) I_{\nu \sigma}(k_{1}) +
\eqno{(6.6)}
$$
$$
+ N_{c}{\rm Im}({\cal D}_{\rho \alpha}(k)) \int \, dk_{1}dk_{2}
\, (S_{k,k_{1},k_{2}}^{\rho \mu \nu} - S_{k,k_{2},k_{1}}^{\rho \nu \mu})
(S_{k,k_{1},k_{2}}^{\ast \alpha \lambda \sigma}
- S_{k,k_{2},k_{1}}^{\ast \alpha \sigma \lambda})I_{\mu \lambda}(k_{1})
I_{\nu \sigma}(k_{2}) \delta (k - k_{1} - k_{2}).
$$

As it is known [23, \,24], in global equilibrium QGP the oscillations of three
typies can be extended: the longitudinal, transverse and
nonphysical (4-D longitudinal) ones. In this connection
we define the Wigner function
$I_{\mu \nu}(k,x)=I_{\mu \nu}$ in the form of expansion
$$
I_{\mu \nu}=P_{\mu \nu}I_{k}^{t} + Q_{\mu \nu}I_{k}^{l} +
\xi D_{\mu \nu}I_{k}^{n} \; , \; I_{k}^{(t,l,n)} \equiv I^{(t,l,n)}(k,x).
\eqno{(6.7)}
$$
The Lorentz matrices in (6.7) are members of the basis [24, \,25]
$$
P_{\mu \nu} (k) = g_{\mu \nu} -
D_{\mu \nu}(k) -
Q_{\mu \nu}(k), \,
Q_{\mu \nu}(k) =
\frac{\bar{u}_{\mu} (k) \bar{u}_{\nu} (k)}{\bar{u}^2(k)}, \,
C_{\mu \nu} (k) =
- \frac{(\bar{u}_{\mu} (k) k_{\nu} +
\bar{u}_{\nu} (k) k_{\mu})}{\sqrt{- 2 k^2 \bar{u}^2(k)}},
$$
$$
D_{\mu \nu} = k_{\mu} k_{\nu}/k^2, \,
\bar{u}_{\mu} (k) = k^2 u_{\mu} - k_{\mu} (k u).
$$
The effective gluon propagator (6.3) can be written in more
conveniet form
$$
{\cal D}_{\mu \nu}(k)= -
P_{\mu \nu}(k) \Delta^t(k) -
Q_{\mu \nu}(k) \Delta^l(k) +
\xi D_{\mu \nu}(k) \Delta^0(k),
\eqno{(6.8)}
$$
where $\Delta^{t,l}(k) = 1/(k^2 - \Pi^{t,l}(k)),
\Pi^t = \frac{1}{2} \Pi^{\mu \nu} P_{\mu \nu},
\Pi^l = \Pi^{\mu \nu} Q_{\mu \nu};
\Delta^0(k) = 1/(( \omega + i \epsilon)^2 - {\bf k}^2)$.
The shift $i \epsilon$ is introduced in $\Delta^0(k)$
to provide the right analytical properties.
At finite temperature, the velocity of plasma introduces
a preferred direction in space-time which breaks manifest Lorentz invariance.
Let us assume that we are in the rest frame of the heat bath,
so that  $u_{\mu}=(1,0,0,0).$

Further derivation of kinetic equation for longitudinal oscillations
involves the same type of manipulations as
in the theory of electromagnetic
plasma, so we can afford to be sketchy.

Now we omit nonlinear terms and anti-Hermitian part of the polarization
tensor in Eq. (3.5). Further substituting the function
$\delta^{ab}Q_{\mu \nu}(k)I_{k}^{l} \delta(k^{\prime} - k)$ instead
of $I_{\mu \nu}^{ab}(k^{\prime},k)$, we lead to the equation
$$
{\rm Re} \, ( \varepsilon^{l}(k)) \, I_{k}^{l} =0.
$$
Here, we use relation
$$
\Delta^{-1 \,l}(k) = k^2 {\varepsilon}^{l}(k),
\eqno{(6.9)}
$$
where
$$
{\varepsilon}^{l}(k) = 1 + \frac{3 \omega_{pl}^2}{{\bf k}^2}
\big[ 1 - F( \frac{\omega}{\vert {\bf k} \vert}) \big] \; , \;
F(x) = \frac{x}{2} \big[ \ln \bigg{\vert} \frac{1 + x}{1 - x} \bigg{\vert}
- i \pi \theta(1 - \vert x \vert) \big] \;
$$
is the longitudinal colour electric permeability and $\omega_{pl}^2 =
\frac{1}{18} g^2 T^2 (N_{f} + 2N_{c})$ is a plasma frequency.
The solution of this equation has the structure
$$
I_{k}^{l}=I_{\bf k}^{l} \delta ( \omega - {\omega}_{\bf k}^{l}) +
I_{- \bf k}^{l} \delta( \omega + {\omega}_{\bf k}^{l}) \; , \;
{\omega}_{\bf k}^{l} >0,
\eqno{(6.10)}
$$
where $I_{\bf k}^{l}$ is a certain function of a wave vector ${\bf k}$
and ${\omega}_{\bf k}^{l} \equiv {\omega}^{l}({\bf k})$ is a frequency
of the longitudinal eigenwaves in QGP.

The equation describing the variation of spectral density of
longitudinal oscillations is obtained from Eq. (6.6) by replacement:
$I_{\mu \nu} \rightarrow Q_{\mu \nu}(k)I_{k}^{l}$, where $I_{k}^{l}$
is defined by (6.10). $\delta$-functions in (6.10)
enable us to remove
integration over frequency and thus we have instead of Eq. (6.6)
$$
\left( k^{2} \frac{\partial {\rm Re} {\varepsilon}^{l}(k)}
{\partial k_{\lambda}} \right)_{\omega = \omega_{\bf k}^{l}}
\frac{\partial I_{\bf k}^{l}}{\partial x^{\lambda}}=
2N_{c}I_{\bf k}^{l} \int \, d{\bf k}_{1} \, I_{{\bf k}_{1}}^{l}
\big( {\rm Im} \,
[( \tilde{\Sigma}_{k,k,k_{1},-k_{1}}^{\mu \nu \lambda \sigma} -
\tilde{\Sigma}_{k,k_{1},k,-k_{1}}^{\mu \nu \sigma \lambda}) +
$$
$$
+ ( \tilde{\Sigma}_{k,k,-k_{1},k_{1}}^{\mu \nu \lambda \sigma} -
\tilde{\Sigma}_{k,-k_{1},k,k_{1}}^{\mu \nu \sigma \lambda})]
Q_{\mu \lambda}(k)Q_{\nu \sigma}(k_{1}) \big)_{\omega= \omega_{\bf k}^{l}, \,
\omega_{1}= \omega_{{\bf k}_{1}}^{l}} +
\eqno{(6.11)}
$$
$$
+ N_{c} \int_{0}^{\infty} d \omega
\int \, d{\bf k}_{1} d{\bf k}_{2} I_{{\bf k}_{1}}^{l}
I_{{\bf k}_2}^{l} (
G_{k,k_1,k_2} + G_{k,-k_1,k_2} + G_{k,k_1,-k_2} + G_{k,-k_1,-k_2}
)_{\omega_{1}= \omega_{{\bf k}_{1}}^{l}, \,
\omega_{2}= \omega_{{\bf k}_{2}}^{l}} \, ,
$$
where
$$
G_{k,k_1,k_2} = {\rm Im}({\cal D}_{\rho \alpha}(k))
(S_{k,k_{1},k_{2}}^{\rho \mu \nu} - S_{k,k_{2},k_{1}}^{\rho \nu \mu})
(S_{k,k_{1},k_{2}}^{\ast \alpha \lambda \sigma}
- S_{k,k_{2},k_{1}}^{\ast \alpha \sigma \lambda})Q_{\mu \lambda}(k_{1})
Q_{\nu \sigma}(k_{2}) \delta(k - k_{1} -k_{2}).
$$

Let us consider the terms in the r.h.s. of
Eq. (6.11). The integral with the function $G_{k,k_{1},k_{2}}$ is different
from zero if the conservation laws are obeyed
$$
{\bf k}={\bf k}_{1} + {\bf k}_{2},
$$
$$
\omega_{\bf k}^{l} = \omega_{{\bf k}_1}^{l} +
\omega_{{\bf k}_2}^{l}.
$$
These conservation laws describe a decay of one longitudinal wave into
two longitudinal waves. However for a dispersion law of the longitudinal oscillations
in QGP, these resonance equations have no solutions,
i.e. this nonlinear process is forbidded. Therefore the integral
with $G_{k,k_{1},k_{2}}$ vanishes. Remaining integrals with $G$-functions
differ from $G_{k, k_1, k_2}$ in that some of the interacting waves are not radiated
but absorped. They also vanish.

The expression
$$
( \tilde{\Sigma}_{k,k,-k_{1},k_{1}}^{\mu \nu \lambda \sigma} -
\tilde{\Sigma}_{k,- k_{1},k,k_{1}}^{\mu \nu \sigma \lambda} ) \,
Q_{\mu \lambda}(k)Q_{\nu \sigma}(k_{1}) \mid_{\omega=
\omega_{\bf k}^{l} , \, \omega_{1}= \omega_{{\bf k}_1}^{l}},
\eqno{(6.12)}
$$
contains the factors
$$
1/(pk + ip_{0} \epsilon) \, , \, 1/(pk_{1} + ip_{0} \epsilon) \, ,
\, 1/(p(k - k_{1}) + ip_{0} \epsilon) .
$$
Imaginary parts of first two factors should be setting equal to
zero, because they are connected with linear Landau damping of
longitudinal waves (which is absent in QGP), and therefore, the imaginary
part of the expression (6.12)
will be defined as
$$
{\rm Im} \frac{1}{p(k - k_{1}) + ip_{0} \epsilon}
\bigg{\vert}_{\omega = \omega_{\bf k}^{l}, \,
\omega_{1}= \omega_{{\bf k}_{1}}^{l}} =
- \frac{i \pi}{p_{0}} \delta (\omega_{\bf k}^{l} - \omega_{{\bf k}_{1}}^{l}
- {\bf v}({\bf k} - {\bf k}_{1})).
$$
It follows that nonlinear term in the r.h.s. of (6.11)
with the function
(6.12) is different from zero if the conservation law is obeyed
$$
\omega_{\bf k}^{l} - \omega_{{\bf k}_{1}}^{l}
- {\bf v}({\bf k} - {\bf k}_{1})=0.
\eqno{(6.13)}
$$
This conservation law describes the process of scattering of plasmon
by the hard thermal particle in QGP.

Let us consider in more detail the term in (6.12)
(see definition (6.5)
with propagator ${\cal D}_{\rho \alpha}(k - k_{1})$. By expansion (6.8)
this propagator represents the nonlinear interaction of
longitudinal waves with longitudinal ones through three types of
intermediate oscillations: the transverse, longitudinal and nonphysical
oscillations depending on a gauge parameter. The term with
$$
\left( P_{\rho \alpha}(k - k_1) \Delta^{t}(k - k_1) \right)_{\omega =
\omega_{\bf k}^{l}, \, \omega_{1} = \omega_{{\bf k}_{1}}^{l}}
$$
in general, describes two
fundamentally different nonlinear processes:
\begin{enumerate}
\item if $( \omega_{\bf k}^{l} - \omega_{{\bf k}_{1}}^{l},
{\bf k} - {\bf k}_{1})$ is a solution of the dispersion equation
$\Delta^{t}(k - k_{1})=0$, then this term describes the fusion process
of two longitudinal oscillations in transverse eigenwave;
\item otherwise, it defines the process of nonlinear scattering of
longitudinal waves in longitudinal those through the transverse virtual
oscillation (for a virtual wave in distinction to the eigenwave,
a frequency $\omega$ and a wave vector ${\bf k}$ are not connected with
each other by the dispersion dependence $\omega \neq \omega({\bf k})$).
\end{enumerate}
The equality $\Delta^{l}(k - k_{1}) \vert_{\omega = \omega^{l}_{\bf k},
\, \omega_1 = \omega^{l}_{{\bf k}_1}} = 0$
does not hold for longitudinal oscillations,
as we see above and therefore, the term
$$
\left( Q_{\rho \alpha}(k - k_{1}) \Delta^l(k - k_1) \right)_{\omega =
\omega_{\bf k}^{l}, \omega_{1} = \omega_{{\bf k}_{1}}^{l}}
$$
defines only the process of scattering of longitudinal waves in longitudinal
those through the longitudinal virtual oscillation.

The contribution of nonphysical intermediate oscillations
$$
\left( \xi \, D_{\rho \alpha}(k - k_1) \Delta^{0}(k - k_1) \right)_{\omega=
\omega_{\bf k}^{l}, \omega_1 = \omega_{{\bf k}_1}^{l}}
$$
will be considered in Sec. 12.

The remaining terms in the Eq. (6.11) with $\tilde{\Sigma}$
are distinguished from above
considered ones by a sign of $k_{1}$, and describe the processes of simultaneous
radiation or absorption by particles of two plasmons.

Summing the preceding and going to the function
$$
W_{\bf k}^{l}= - \left( \omega k^{2} \frac{\partial {\rm Re}
\varepsilon^{l}(k)}{\partial \omega} \right)_{\omega =
\omega_{\bf k}^{l}} I_{{\bf k}}^{l} \, ,
$$
having the physical meaning of spectral density of
longitudinal oscillations energy, we find from (6.11) the required kinetic
equation for longitudinal waves in QGP
$$
\frac{\partial W_{\bf k}^{l}}{\partial t} +
{\bf V}_{\bf k}^{l} \, \frac{\partial W_{\bf k}^{l}}
{\partial {\bf x}} = - \hat{\gamma} \{
( \frac{W_{\bf k}^{l}}{\omega_{\bf k}^{l}})  \} \, W_{\bf k}^{l},
\eqno{(6.14)}
$$
where
$$
{\bf V}_{\bf k}^{l} = \frac{\partial \omega_{\bf k}^{l}}
{\partial {\bf k}} =
- \Big[ \big( \frac{\partial {\rm Re} \,
\varepsilon^{l}(k)}{\partial {\bf k}} \big)
\Big/ \big( \frac{\partial {\rm Re} \, \varepsilon^{l}(k)}
{\partial \omega} \big) \Big]_{\omega = \omega_{\bf k}^{l}}
$$
is the group velocity of longitudinal oscillations and
$$
\hat{\gamma} \{ \Big( \frac{W_{\bf k}^{l}}{\omega_{\bf k}^{l}} \Big) \}
\equiv \gamma^{l}({\bf k}) = 2N_{c} \int \, d{\bf k}_{1}
\Big( \frac{W_{{\bf k}_{1}}^{l}}{\omega_{{\bf k}_{1}}^{l}} \Big)
\bigg[ \frac{1}{k^{2}k^{2}_{1}}
\Big( \frac{\partial {\rm Re} \, \varepsilon^{l}(k)}
{\partial \omega} \Big)^{-1}
\Big( \frac{\partial {\rm Re} \, \varepsilon^{l}(k_{1})}
{\partial \omega_{1}} \Big)^{-1}
\eqno{(6.15)}
$$
$$
{\rm Im}[( \tilde{\Sigma}_{k,k,-k_{1},k_{1}}^{\mu \nu \lambda \sigma} -
\tilde{\Sigma}_{k,-k_{1},k,k_{1}}^{\mu \nu \sigma \lambda}) +
( \tilde{\Sigma}_{k,k,k_{1},-k_{1}}^{\mu \nu \lambda \sigma} -
\tilde{\Sigma}_{k,k_{1},k,-k_{1}}^{\mu \nu \sigma \lambda})]
Q_{\mu \lambda}(k)Q_{\nu \sigma}(k_{1}) \bigg]_{\omega = \omega_{\bf k}^{l},
\, \omega_{1}= \omega_{{\bf k}_{1}}^{l}}
$$
presents the damping rate caused by nonlinear effects and being the linear
functional of spectral density of energy.

One can write (6.14) in more convenient form
if the spectral density of number of longitudinal oscillations is
represented as
$$
N_{\bf k}^{l} = W_{\bf k}^{l}/ \omega_{\bf k}^{l}.
$$
It fulfils the role of distribution function of plasmons number.
Then instead of (6.14) we have
$$
\frac{{\rm d} N_{\bf k}^{l}}{{\rm d}t} \equiv
\frac{\partial N_{\bf k}^{l}}{\partial t} +
{\bf V}_{\bf k}^{l} \, \frac{\partial N_{\bf k}^{l}}
{\partial {\bf x}} = - \hat{\gamma} \, \{
N_{\bf k}^{l} \} \, N_{\bf k}^{l} \, .
\eqno{(6.16)}
$$
\vspace{1cm}\\
{\bf 7. HTL-AMPLITUDES. WARD IDENTITIES}\\

Befor proceeding to an queshion on a gauge dependence of obtained
nonlinear Landau damping rate (6.15), we rewrite derived expression in the terms
of HTL-amplitudes [1, \,12]. This makes possible to extend a procedure of a
gauge invariance proof of the damping rate for QGP collective excitations
in quantum theory [1] to our case.

We present the integration measure ${\rm d}^4p$ as
${\rm d} p^0 \vert{\bf p} \vert^2
{\rm d} \vert {\bf p} \vert {\rm d} \Omega$, where ${\rm d} \Omega$
is the angular measure. Using the definition of equilibrium
distributions (2.17)
(for $\mu=0$) and taking into account
$$
\int \limits_{- \infty}^{+ \infty} \vert {\bf p} \vert^2 \,
{\rm d} \vert {\bf p} \vert
\int \limits_{- \infty}^{+ \infty} p_0 {\rm d} p_0
\frac{d {\cal N}_{eq}(p_0)}{d p_0} = - \frac{3}{4 \pi}
\left( \frac{\omega_{pl}}{g} \right)^2,
$$
we perform the integral over ${\rm d} p_0$
and the radial integral over ${\rm d} \vert {\bf p} \vert$ in the expressions
for $S^{(II)}$-function (5.5) and $\Sigma^{(II)}$-function (5.10).
Futher we rewrite the expression (6.15) as follows
$$
\gamma^{l}({\bf k}) =
-2 g^2 N_{c} \int {\rm d} {\bf k}_1 \Big( \frac{W^{l}_{{\bf k}_1}}{\omega^{l}_
{{\bf k}_1}} \Big)
\Big[ \frac{1}{{k^2}{k^2_1}}
\Big( \frac{\partial {\rm Re} \, \varepsilon^{l}(k)}{\partial \omega} \Big)^{-1}
\Big( \frac{\partial {\rm Re} \, \varepsilon^{l}(k_1)}{\partial \omega_1}
\Big)^{-1} \Big]_{on-shell} {\rm Im} \, \tilde{T} ({\bf k}, {\bf k}_1),
\eqno{(7.1)}
$$
where
$$
\tilde{T}({\bf k},{\bf k}_1) \equiv
\eqno{(7.2)}
$$
$$
\equiv
\{ \delta \Gamma^{\mu \nu \lambda \sigma}(k,k_1,-k,-k_1) \, -
\,^{\ast} \Gamma^{\mu \nu \rho}(k,-k_1,-k+k_1){\cal D}_{\rho \alpha}
(k - k_1) \,^{\ast} \Gamma^{\alpha \lambda \sigma}(k - k_1,-k,k_1) \, -
$$
$$
- \,^{\ast} \Gamma^{\mu \nu \rho}(k,k_1,-k-k_1){\cal D}_{\rho \alpha}
(k + k_1) \,^{\ast} \Gamma^{\alpha \lambda \sigma}(k + k_1,-k,-k_1) \}
Q_{\mu \lambda}(k)Q_{\nu \sigma}(k_1) \vert_
{\omega = \omega^{l}_{\bf k}, \;
\omega_1 = \omega^{l}_{{\bf k}_1}}
$$
and
$$
\delta \Gamma^{\mu \nu \lambda \sigma}(k, k_1, k_2,k_3)=
3 \, \omega^2_{pl} \int \, \frac{{\rm d} \Omega}{4 \pi} \,
\frac{v^{\mu}v^{\nu}v^{\lambda}v^{\sigma}}{vk + i \epsilon} \, \Big[
\frac{1}{v(k + k_1) + i \epsilon} \Big( \frac{\omega_{2}}{vk_2 - i \epsilon} -
$$
$$
- \frac{\omega_3}{vk_3 - i \epsilon} \Big)
- \frac{1}{v(k + k_3) + i \epsilon} \Big( \frac{\omega_{1}}{vk_1 - i \epsilon} -
\frac{\omega_2}{vk_2 - i \epsilon} \Big) \Big] \; , \; (v^{\mu} = (1,{\bf {\bf v}}))
\eqno{(7.3)}
$$
present HTL-corrections to the bare four-gluon vertex $[1, 12]$.
$$
\,^{\ast} \Gamma^{\mu \nu \rho}(k,k_1,k_2) \equiv
\Gamma^{\mu \nu \rho}(k,k_1,k_2) +
\delta \Gamma^{\mu \nu \rho}(k,k_1,k_2)
\eqno{(7.4)}
$$
is the effective three-gluon vertex [1, 12],
that is a sum of the bare three-gluon vertex
$$
\Gamma^{\mu \nu \rho}(k,k_1,k_2) =
g^{\mu \nu} (k - k_1)^{\rho} + g^{\nu \rho} (k_1 - k_2)^{\mu}+
g^{\mu \rho} (k_2 - k)^{\nu}
\eqno{(7.5)}
$$
and corresponding HTL-correction
$$
\delta \Gamma^{\mu \nu \rho}(k,k_1,k_2) =
3 \, \omega^2_{pl} \int \, \frac{{\rm d} \Omega}{4 \pi} \,
\frac{v^{\mu}v^{\nu}v^{\rho}}{vk + i \epsilon} \,
\Big( \frac{\omega_2}{vk_2 - i \epsilon} -
\frac{\omega_1}{vk_1 - i \epsilon} \Big).
\eqno{(7.6)}
$$
The polarization tensor in these notations takes the form
$$
\Pi^{\mu \nu}(k) = 3 \, \omega_{pl}^{2}
\left( u^{\mu}u^{\nu} - \omega \int \, \frac{{\rm d} \Omega}{4 \pi}
\frac{v^{\mu}v^{\nu}}{vk + i \epsilon} \right).
$$

For writing the expression (7.1), for example, in the
temporal gauge $A_{0}^a=0$, it is sufficiently to replace the
projection operators $Q_{\mu \lambda}(k)$
and $Q_{\nu \sigma}(k_1)$ by
$$
\tilde{Q}_{\mu \lambda}(k) = Q_{\mu \lambda}(k) +
\frac{\sqrt{-2k^2 {\bar{u}}^2}}{k^2(ku)}C_{\mu \lambda}(k) +
\frac{\bar{u}^2(k)}{k^2(ku)^2}D_{\mu \lambda}(k)=
\frac{\tilde{u}_{\mu}(k) \tilde{u}_{\lambda}(k)}{\bar{u}^2(k)},
$$
$$
\tilde{u}_{\mu}(k) \equiv \frac{k^2}{(ku)}(k_{\mu} - u_{\mu}(ku)),
$$
(similarly for $Q_{\nu \sigma}(k_1)$), and the propagator
(6.8) by
$$
\tilde{\cal D}_{\rho \alpha}(k) = {\cal D}_{\rho \alpha}(k) -
\Big( \frac{\sqrt{-2k^2 {\bar{u}}^2}}{k^2(ku)}C_{\rho \alpha}(k) +
\frac{\bar{u}^2(k)}{k^2(ku)^2}D_{\rho \alpha}(k) \Big) \Delta^{l}(k) -
$$
$$
- \xi D_{\rho \alpha}(k) \Delta^{0}(k) -
\xi_{0} \frac{k^2}{(ku)^2} D_{\rho \alpha}(k),
\eqno{(7.7)}
$$
where $\xi_{0}$ is a gauge parameter in the temporal gauge.

To establish the gauge invariance of nonlinear Landau damping rate
$\gamma^{l}({\bf k})$ there is a need to show that expression
${\rm Im} \, \tilde{T}({\bf k},{\bf k}_1)$, where function
$\tilde{T}({\bf k},{\bf k}_1)$ defined by (7.2)
in a covariant gauge equals to similar
expression in the temporal gauge.

We prove a gauge invariance for more general expression that in a covariant
gauge has the form
$$
\,^{\ast} \tilde{\Gamma}(k,-k_2,k_1,-k_3) \equiv
\eqno{(7.8)}
$$
$$
\equiv \{ \,^{\ast} \Gamma^{\mu \sigma \lambda \nu}(k,-k_2,k_1,-k_3) -
{\cal D}_{\rho \alpha}(-k_1 +k_2) \,^{\ast} \Gamma^{\mu \nu \rho}
(k,-k_3,k_1 - k_2) \,^{\ast} \Gamma^{\alpha \lambda \sigma}(-k_1 + k_2,k_1,-k_2) -
$$
$$
- {\cal D}_{\rho \alpha}(-k_1 +k_3) \,^{\ast} \Gamma^{\mu \sigma \rho}(k,-k_2,k_1 - k_3)
\,^{\ast} \Gamma^{\alpha \lambda \nu}(-k_1 + k_3,k_1,-k_3) \}
\bar{u}_{\mu}(k) \bar{u}_{\lambda}(k_1)
\bar{u}_{\nu}(k_3) \bar{u}_{\sigma}(k_2)
\vert_{on-shell}.
$$
Here, $k +k_1 = k_2 + k_3$. The effective vertex
$\,^{\ast} \Gamma^{\mu \sigma \lambda \nu}$
are formed by adding the hard thermal loop (7.3) to the
bare four-gluon vertex
$$
\Gamma^{\mu \sigma \lambda \nu} =
2g^{\mu \sigma}g^{\lambda \nu} - g^{\mu \nu}g^{\sigma \lambda} -
g^{\mu \lambda}g^{\nu \sigma}.
$$
The association of the expression (7.2) with (7.8)
is given by
\[
{\rm Im} \, \tilde{T}({\bf k},{\bf k}_1) =
\frac{1}{\bar{u}^2(k) \bar{u}^2(k_1)}
{\rm Im} \,^{\ast}\tilde{\Gamma}(k,-k_2,k_1,-k_3) \vert_{k_1 = -k,
\; k_2 = -k_1, \; k_3 = k_1}.
\]
As will be shown in a separate publication [26],
the expression (7.8) is associated with probability of
plasmon-plasmon scattering.

Similar expression (7.8) in the temporal gauge is obtained with
replacements $\bar{u}_{\mu}(k) \rightarrow  \tilde{u}_{\mu}(k)$
and the propagator (6.8) by (7.7).

The gauge invariance proof is based on using the identities, analogous
the effective Ward identities in hot gauge theories [1]. It can be
shown that the following equalities hold
$$
k_{\mu} \,^{\ast} \Gamma^{\mu \nu \lambda \sigma}(k,k_1,k_2,k_3) =
\,^{\ast} \Gamma^{\nu \lambda \sigma}(k_1,k_2,k+k_3) -
\,^{\ast} \Gamma^{\nu \lambda \sigma}(k+k_1,k_2,k_3),
$$
$$
k_{1 \nu} \,^{\ast} \Gamma^{\mu \nu \lambda \sigma}(k,k_1,k_2,k_3) =
\,^{\ast} \Gamma^{\mu \lambda \sigma}(k+k_1,k_2,k_3) -
\,^{\ast} \Gamma^{\mu \lambda \sigma}(k,k_1+k_2,k_3),
\eqno{(7.9)}
$$
(similar contractions with $k_{2 \lambda}, k_{3 \sigma}$),
$$
k_{\mu} \,^{\ast} \Gamma^{\mu \nu \rho}(k,k_1,k_2) =
{\cal D}^{-1 \, \nu \rho}(-k_1) - {\cal D}^{-1 \, \nu \rho}(-k_2),
\eqno{(7.10)}
$$
(similar contractions with $k_{1 \nu}, k_{2 \rho}$).

Here ${\cal D}^{-1 \, \mu \nu}(k) = P^{\mu \nu }(k) \Delta^{-1 \, t}(k) +
Q^{\mu \nu}(k) \Delta^{-1 \, l}(k)$ is the inverse propagator for which
the following useful relation holds
$$
{\cal D}_{\rho \alpha} {\cal D}^{-1 \, \alpha \lambda}(k) =
\delta^{\lambda}_{\rho} - \frac{k_{\rho}k^{\lambda}}{k^2}.
\eqno{(7.11)}
$$
\vspace{1cm}\\
{\bf 8. THE GAUGE INVARIANCE}\\

Our proof of gauge invariance is reduced to contraction
of the projectors $Q$ and $\tilde{Q}$ with resummed three- and
four-gluon vertices and the use of the Ward identities (7.9), (7.10).
At the beginning we consider the term with a gauge parameter for the propagator
in a covariant gauge.

By using the Ward identities (7.10), we have
$$
\xi D_{\rho \alpha}(k_1 - k_2) \Delta^{0}(k_1 - k_2)
\,^{\ast}\Gamma^{\mu \nu \rho}(k_1, -k_3,k_1 - k_2)
\,^{\ast}\Gamma^{\alpha \lambda \sigma}(-k_1+k_2,k_1,-k_2) =
$$
$$
= \xi (\Delta^{0}(k_1 - k_2))^2 ({\cal D}^{-1 \, \mu \sigma}(k)
- {\cal D}^{-1 \, \mu \sigma}(k_3))
({\cal D}^{-1 \, \lambda \nu}(k_2) - {\cal D}^{-1 \, \lambda \nu}(-k_1)).
\eqno{(8.1)}
$$
Further this expression is contracted with
$\bar{u}_{\mu}(k) \bar{u}_{\nu}(k_3)$.
It is easily shown that it vanish whether because
${\cal D}^{- 1 \, \mu \nu} (k)$ is transverse, or
by the definition of the mass-shall condition, i.e.
$$
k_{\mu}{\cal D}^{-1 \, \mu \nu}(k) = 0, \;
{\cal D}^{-1 \, 0 \nu}(k) \vert_{\omega= \omega^{l}_{\bf k}} = 0.
\eqno{(8.2)}
$$
Similar statement holds in the temporal gauge also.

The gauge-dependent parts in the above calculation drops out
$\gamma^l({\bf k})$, since they are multiplied by the mass-shell factor.
These factors are proportional to $(\omega - \omega_{\bf k}^{l})$.
However, in a quantum case Baier, Kunstatter, and Schiff [27]
observed that naive calculation in covariant gauge appears to violate
this consideration. Mass-shell factor is multiplied
by the integral involving
a power infrared divergence which is cutoff exactly on the scale
$(\omega - \omega_{\bf k}^{l}) \sim g^2 T$.
By this means the gauge-dependent
part yields a finite contribution to the gluon damping rate. This problem
is considered for damping rate of Fermi-excitations in QGP, also [28, \,
5, \,29].

It can be shown that in our case, the integral preceding the
mass-shell conditions, diverges for lower limit also
and thus the similar problem is arised:
does (8.1) yield a finite, the gauge-dependent
contribution to the nonlinear Landau damping rate?
In Sec. 12 we provide the answer to the queshion.

Now we consider the remaining terms in (7.8). We calculate
the contraction with effective four-gluon vertex $ \,^{\ast} \Gamma_4$.
Slightly cumbersome, but not complicated computations by using the effective
Ward identities (7.9), (7.10) and relations (7.11),
(8.2) lead to the following expression
$$
\,^{\ast} \Gamma^{\mu \sigma \lambda \nu}(k,-k_2,k_1,-k_3)
\bar{u}_{\mu}(k) \bar{u}_{\lambda}(k_1) \bar{u}_{\nu}(k_3)
\bar{u}_{\sigma}(k_2) \vert_{on-shell} =
$$
$$
= k^2 k^2_1 k^2_2 k^2_3 \,^{\ast} \Gamma^{0000}(k,-k_2,k_1,-k_3) +
\Xi (k,-k_2,k_1,-k_3),
\eqno{(8.3)}
$$
where
$$
\Xi (k,-k_2,k_1,-k_3)=
\{ (k_1^2k_2^2[ \omega k_3^2 + \omega_3 k^2] \,^{\ast} \Gamma^{000}(k-k_3,k_1,-k_2)-
$$
$$
- \omega \omega_3 k_1^2 k_2^2 k_{3 \nu} \,^{\ast} \Gamma^{\nu 00}(k-k_3,k_1,-k_2))
+(k \leftrightarrow k_3) + (k_1 \leftrightarrow k) +
(k_1 \leftrightarrow k, k_2 \leftrightarrow k_3) \} +
$$
$$
+ \{ (( \omega k_3^2 + \omega_3k^2)( \omega_2 k_1^2 + \omega_1 k_2^2)
{\cal D}^{-1 \,00}(-k_1 + k_2) - \omega_1 \omega_2( \omega k_3^2 + \omega_3 k^2)
k_{1 \lambda}{\cal D}^{-1 \,0 \lambda}(-k_1 + k_2) -
$$
$$
- \omega \omega_3( \omega_2 k_1^2 + \omega_1 k_2^2) k_{3 \nu}{\cal D}^{-1
\,0 \nu}(-k_1 + k_2) + \omega \omega_1 \omega_2 \omega_3
k_{1 \lambda}k_{3 \nu}{\cal D}^{-1 \nu \lambda}(-k_1 + k_2)) +
(k_2 \leftrightarrow k_3) \}.
$$

Further, we calculate the contractions with the terms
containing $\,^{\ast} \Gamma_3$ in (7.8).
Here, we are led to the expression
$$
\{ {\cal D}_{\rho \alpha}(-k_1 + k_2)
\,^{\ast} \Gamma^{\mu \nu \rho}(k,-k_3,k_1 - k_2)
\,^{\ast} \Gamma^{\alpha \lambda \sigma}(-k_1 + k_2,k_1,- k_2) +
(k_2 \leftrightarrow k_3) \}
\bar{u}_{\mu}(k) \bar{u}_{\lambda}(k_1)
$$
$$
\bar{u}_{\nu}(k_3) \bar{u}_{\sigma}(k_2) \vert_{on-shell} =
\eqno{(8.4)}
$$
$$
= k^2 k^2_1 k^2_2 k_3^2
\{ {\cal D}_{\rho \alpha}(-k_1 + k_2)
\,^{\ast} \Gamma^{0 0 \rho}(k,-k_3,k_1 - k_2)
\,^{\ast} \Gamma^{\alpha 0 0}(-k_1 + k_2,k_1,- k_2) +
(k_2 \leftrightarrow k_3) \}
$$
$$
+ \Xi(k,-k_2,k_1,-k_3).
$$
Note that the function $\Xi$ in the r.h.s. of (8.4)
is identical to that in (8.3). Subtracting (8.4) from
(8.3), we arrive at desired expression
$$
\,^{\ast} \tilde{\Gamma}(k,-k_2,k_1,-k_3) = k^2 k_1^2 k_2^2 k_3^2 \{
\,^{\ast} \Gamma^{0000}(k,-k_2,k_1,-k_3) -
{\cal D}_{\rho \alpha}(k - k_3) \,^{\ast} \Gamma^{00 \rho}(k,-k_3,k_1 - k_2)
$$
$$
\,^{\ast} \Gamma^{\alpha 00}(-k_1+k_2,k_1, - k_2) -
{\cal D}_{\rho \alpha}(k - k_2) \,^{\ast} \Gamma^{00 \rho}(k,-k_2,k_1 - k_3)
\,^{\ast} \Gamma^{\alpha 00}(-k_1 + k_3,k_1,-k_3) \}.
\eqno{(8.5)}
$$

Now we consider the structure of $\,^{\ast} \tilde{\Gamma}$
in the temporal gauge. For this purpose we replace  $\bar{u}_{\mu}$
by $\tilde{u}_{\mu}$ in (7.8) and the propagator in the covariant
gauge by the propagator in the temporal gauge (7.7).
The contraction with effective four-gluon vertex leads to
$$
\,^{\ast} \Gamma^{\mu \sigma \lambda \nu}(k,-k_2,k_1,-k_3)
\tilde{u}_{\mu}(k) \tilde{u}_{\lambda}(k_1) \tilde{u}_{\nu}(k_3)
\tilde{u}_{\sigma}(k_2) \vert_{on-shell} =
$$
$$
= k^2 k^2_1 k^2_2 k^2_3 \{ \,^{\ast} \Gamma^{0000}(k,-k_2,k_1,-k_3) +
\tilde{\Xi} (k,-k_2,k_1,-k_3) \},
\eqno{(8.6)}
$$
where
$$
\tilde{\Xi} (k,-k_2,k_1,-k_3)=
\{ ( \omega_1 \omega_2( \omega + \omega_3) \,^{\ast} \Gamma^{000}(k-k_3,k_1,-k_2)-
$$
$$
-\omega_1 \omega_2 k_{3 \nu} \,^{\ast} \Gamma^{\nu 00}(k-k_3,k_1,-k_2))+
(k \leftrightarrow k_3) + (k_1 \leftrightarrow k) +
(k_1 \leftrightarrow k, k_2 \leftrightarrow k_3) \} +
$$
$$
+ \{ (( \omega_1 + \omega_2)( \omega + \omega_3)
{\cal D}^{-1 \,00}(-k_1 + k_2) - ( \omega + \omega_3)
k_{1 \lambda}{\cal D}^{-1 \,0 \lambda}(-k_1 + k_2) -
$$
$$
- ( \omega_1 + \omega_2) k_{3 \nu}{\cal D}^{-1 \,
0 \nu}(-k_1 + k_2) +
k_{1 \lambda}k_{3 \nu}{\cal D}^{-1 \nu \lambda}(-k_1 + k_2)) +
(k_2 \leftrightarrow k_3) \}.
$$

Contracting with the terms containing $\,^{\ast} \Gamma_3$, yields
$$
\{ {\tilde{\cal D}}_{\rho \alpha}(-k_1 + k_2)
\,^{\ast} \Gamma^{\mu \nu \rho}(k,-k_3,k_1 - k_2)
\,^{\ast} \Gamma^{\alpha \lambda \sigma}(-k_1 + k_2,k_1,- k_2) +
(k_2 \leftrightarrow k_3) \}
\tilde{u}_{\mu}(k) \tilde{u}_{\lambda}(k_1)
$$
$$
\tilde{u}_{\nu}(k_3) \tilde{u}_{\sigma}(k_2) \vert_{on-shell} =
\eqno{(8.7)}
$$
$$
= k^2 k^2_1 k^2_2 k_3^2
\{ {\tilde{\cal D}}_{\rho \alpha}(-k_1 + k_2)
\,^{\ast} \Gamma^{0 0 \rho}(k,-k_3,k_1 - k_2)
\,^{\ast} \Gamma^{\alpha 0 0}(-k_1 + k_2,k_1,- k_2) +
(k_2 \leftrightarrow k_3)+
$$
$$
+ \tilde{\Xi}(k,-k_2,k_1,-k_3) \}.
$$
Subtracting (8.7) from (8.6), we are led to similar
expession (8.5). Thus, we have shown that at least in a class of the
covariant and temporal gauges, the nonlinear Landau damping rate (7.1)
(exactly, its piece, independent of a gauge parameter)
is a gauge-invariant.
\vspace{1cm}\\
{\bf 9. THE PHYSICAL MECHANISM OF NONLINEAR SCATTERING
OF WAVES}\\

In Sec. 7 the expression of the nonlinear Landau damping rate
$\gamma^{l}({\bf k})$ for Bose excitations in a quark-gluon plasma was
obtained. Now we transform it to the form, in which each term in
$\gamma^{l} ({\bf k})$ has direct physical relevance.
The first transformation of this type has been proposed by
Tsytovich for
Abelian plasma [21]. In [20] it taken into account the contribution
with longitudinal virtual wave only.
In order to satisfy gauge invariance, we extend the transformation
of similar type to the case with the transverse virtual wave.
Since the nonlinear Landau damping rate is independent of
the choice of gauge, we choose the temporal gauge for simplicity.

As was mentioned in Sec. 6, the expression for $\gamma^l({\bf k})$
contains the contributions of two different processes. The first is associated
with absorbtion of a plasmon by QGP particles with frequency $\omega$
and a wave vector ${\bf k}$ with its consequent radiation with frequency
$\omega_1$ and a wave vector ${\bf k}_1$.
It defined by the second term in the r.h.s. of (7.2).
The frequency and wave vectors of a incident plasmon and a recoil plasmon
satisfy the conservation law (6.13).
The second process represents simultaneous radiation (or absorbtion) of two
plasmons with frequency $\omega, \, \omega_1$ and wave vectors
${\bf k},{\bf k}_1$ satisfying the conservation law
$$
\omega^l_{\bf k} + \omega^l_{{\bf k}_1} - {\bf v}({\bf k} +{\bf k}_1)=0,
\eqno{(9.1)}
$$
and it defined by third term in (7.2).
In contrast to previous scattering process, this process not conserves
the plasmons number and its contribution are not important to the order of
interest.
The first term in (7.2), associated with HTL-correction to
four-gluon vertex, contains both processes.
Futher, we have taken into account the terms with (6.13) and
we drop the terms which contain a $\delta$-function of (9.1).

Now we consider the expression with $\delta \Gamma_4$.
With regard to above-mentioned and by the definition (7.3),
the contribution of a given term
to $\gamma^l({\bf k})$ can be represented as
$$
3(\omega_{pl}/g)^2
\int \, (\omega^l_{\bf k} - \omega^l_{{\bf k}_1})
{\it w}_{\bf v}^{c}({\bf k}, {\bf k}_{1})
\left( \frac{W_{{\bf k}_1}^{l}}
{\omega^l_{{\bf k}_1}} \right) \frac{{\rm d} \Omega}{4 \pi}
\, {\rm d}{\bf k}_{1},
\eqno{(9.2)}
$$
where
$$
{\it w}_{\bf v}^{c}({\bf k}, {\bf k}_{1}) =
$$
$$
= 2 \pi N_{c}
\bigg( \frac{\partial {\rm Re} \, \varepsilon^{l}(k)}{\partial \omega}
\bigg)_{\omega = \omega_{\bf k}^{l}}^{-1}
\bigg( \frac{\partial {\rm Re} \, \varepsilon^{l}(k_{1})}{\partial \omega_{1}}
\bigg)_{\omega_{1} = \omega_{\bf k_1}^{l}}^{-1}
\delta ( \omega_{\bf k}^{l} - \omega_{{\bf k}_{1}}^{l} -
{\bf v}( {\bf k} - {\bf k}_{1}))
\vert {\cal M}^{c}({\bf k},{\bf k}_{1}) \vert^{2},
\eqno{(9.3)}
$$
$$
{\cal M}^{c}({\bf k},{\bf k}_{1}) \equiv
\frac{g^{2}}{ \vert {\bf k} \vert \vert {\bf k}_{1} \vert} \,
\frac{1}{\omega^l_{\bf k} \omega^l_{{\bf k}_1}}
\, \frac{({\bf k}{\bf v})({{\bf k}_{1}}{\bf v})}
{\omega_{\bf k}^{l} - ({\bf k}{\bf v})}.
\eqno{(9.4)}
$$

To clear up the physical origin of contribution (9.2),
it is convenient to compare it with corresponding contribution
in the theory of electromagnetic plasma.
In this case as shown in [21]
this contribution describes the normal Thomson scattering
of a wave by particles: a wave with the original frequency $\omega_{\bf k}
^{l}$ is set in oscillatory particle motion
of plasma and
oscillating particle
radiates a wave with modified frequency
$\omega_{{\bf k}_{1}}^{l}$. The corresponding function
${\it w}_{\bf v}^{c}({\bf k},{\bf k}_{1})$ presents the probability of
Thomson scattering. As it was shown above in quark-gluon plasma for a soft
long-wavelength excitations all Abelian contributions is at most $g \ln g$
times the non-Abelian ones and the basic scattering mechanism here, is
essentially another (our discussion is schematic, the details of a
computation displayed in Appendix A).

To show this mechanism we use the classical pattern of QGP
description [8], in which the particles states are characterized besides
position and momentum, by the color vector (non-Abelian charge) ${\rm Q}= (
{\rm Q}^{a}), a=1, \ldots,
N_{c}^{2} - 1 $ also. As was shown by Heinz [8] there is an close
connection between the classical kinetic equations and semiclassical
ones (2.4). Therefore in this case the use of classical notions is
justified.

Let the field acting on a colour particle in QGP represents
a bundle of longitudinal plane waves
$$
\tilde{A}_{\mu}^{a}(x) = - \int \, [( \omega^2/k^2)
Q_{\mu \nu}(k)A_{\bf k}^{a \nu}]_{\omega=
\omega_{\bf k}^{l}} {\rm e}^{i {\bf k}{\bf x} - i \omega_{\bf k}^{l}t}
{\rm d} {\bf k}.
\eqno{(9.5)}
$$
The particle motion in this wave field is described by the system of
equations
$$
{\rm m} \frac{{\rm d}^{2}x^{\mu}}{{\rm d} \tau^{2}} =
g {\rm Q}^{a} \tilde{F}^{a \mu \nu} \frac{{\rm d}x_{\nu}}{{\rm d} \tau},
\eqno{(9.6)}
$$
$$
\frac{{\rm d}{\rm Q}^{a}}{{\rm d} \tau}=
- gf^{abc} \frac{{\rm d}x^{\mu}}{{\rm d} \tau} \tilde{A}_{\mu}^{b}
{\rm Q}^{c}.
\eqno{(9.7)}
$$
Here, $\tau$ is a proper time of a particle. Eq. (9.7) is
familiar Wong equation [30].
The system (9.6), (9.7) is solved
by the approximation scheme method - the ${\it weak \;field \;expansion}$.
A zeroth approximation describes uniform restlinear motion,
and the next one - constrained charge oscillations in the field
(9.5). With a knowledge of the motion law of a charge, the radiation intensity
by it longitudinal waves can be defined. In this case Eq.
(9.6) defines the Abelian contribution to radiation, whereas
(9.7) - non-Abelian
one and interference of these two contributions equals zero. The scattering
probability computed by this means, based on Eq. (9.7) is
coincident with obtained above (9.3).

In this manner the contribution of (9.2) to $\gamma^{l}({\bf k})$
are caused by not the spatially oscillations of a colour particle,
as it occurs in
electromagnetic plasma, but it induced by a precession of a colour
vector ${\rm Q}$ of a particle in field of a longitudinal wave (9.5)
(${\rm Q}^a{\rm Q}^a = {\rm const}$ by Eq. (9.7).

Let us consider now more complicated terms in (7.1) associated
with $\,^{\ast} \Gamma_3$-functions. By the definition of three-gluon
vertex the following equality is obeyed
$$
\,^{\ast} \Gamma^{\mu \nu \rho}(k,-k_1,-k_2)= -
\,^{\ast} \Gamma^{\rho \mu \nu}(k_2,-k,k_1),
$$
(hereafter, $k_2 \equiv k-k_1$). Using this relation, the contribution
of the term with longitudinal virtual wave to $\gamma^l({\bf k})$
can be presented as
$$
-2g^2N_{c} \int \, {\rm } {\rm d}{{\bf k}_{1}}
\left( \frac{W_{{\bf k}_{1}}^{l}}{\omega^l_{{\bf k}_1}} \right)
\frac{1}{{\bf k}^{2}{\bf k}_{1}^{2}{\bf k}_{2}^{2}}
\Big( \frac{\partial {\rm Re} \, \varepsilon^{l}(k)}{\partial \omega}
\Big)_{\omega = \omega_{\bf k}^{l}}^{-1}
\Big( \frac{\partial {\rm Re} \, \varepsilon^{l}(k_{1})}{\partial \omega_1}
\Big)_{\omega_{1} = \omega_{{\bf k}_{1}}^{l}}^{-1}
$$
$$
\Big[ \frac{1}{( \omega \omega_1 \omega_2)^2} {\rm Im} \,
(k_2^2 \Delta^l(k_2)( \,^{\ast} \Gamma^{ijl}(k,-k_1,-k_2)k^ik^j_1k^l_2)^2)
\Big]_{\omega = \omega_{\bf k}^{l}, \, \omega_{1} = \omega_{{\bf k}_{1}}^{l}}.
$$
Futher we use the relation
$$
{\rm Im}  (k_2^2 \Delta^l(k_2)( \,^{\ast} \Gamma^{ijl}k^ik^j_1k^l_2)^2)=
- {\rm Im} \,(k_2^2 \Delta^l(k_2))^{-1}
\vert k_2^2 \Delta^l(k_2)( \,^{\ast} \Gamma^{ijl}k^ik^j_1k^l_2) \vert^2 +
$$
$$
+2( {\rm Im} \, \,^{\ast} \Gamma^{ijl}k^ik^j_1k^l_2)
{\rm Re}( k_2^2 \Delta^l(k_2)( \,^{\ast} \Gamma^{ijl}k^ik^j_1k^l_2)).
\eqno{(9.8)}
$$
Here, we drop the momentum dependence on vertices.
Taking into consideration the equality
\[
{\rm Im} ( \, k_2^2 \Delta^l(k_2))^{-1} =
3 \pi \omega^2_{pl} \frac{\omega_2}{{\bf k}_2^2} \int \,
\frac{{\rm d} \Omega}{4 \pi} \, \delta(vk_{2}) \, ,
\]
one can write contribution of the first term in the r.h.s. of
(9.8) to the nonlinear damping rate in the form
$$
3( \omega_{pl}/g)^2 \int ( \omega^l_{\bf k} - \omega^l_{{\bf k}_1})
\,{\it w}_{\bf v}^{\parallel}({\bf k}, {\bf k}_{1})
\left( \frac{W_{{\bf k}_1}^{l}}{\omega^l_{{\bf k}_1}} \right)
\frac{{\rm d} \Omega}{4 \pi} \, {\rm d}{\bf k}_{1},
$$
where
$$
{\it w}_{\bf v}^{\parallel}({\bf k}, {\bf k}_{1}) =
$$
$$
= 2 \pi N_{c}
\bigg( \frac{\partial {\rm Re} \, \varepsilon^{l}(k)}{\partial \omega}
\bigg)_{\omega = \omega_{\bf k}^{l}}^{-1}
\bigg( \frac{\partial {\rm Re} \, \varepsilon^{l}(k_{1})}{\partial \omega_{1}}
\bigg)_{\omega_{1} = \omega_{\bf k_1}^{l}}^{-1}
\delta ( \omega_{\bf k}^{l} - \omega_{{\bf k}_{1}}^{l} -
{\bf v}( {\bf k} - {\bf k}_{1}))
\vert {\cal M}^{\parallel}({\bf k},{\bf k}_{1}) \vert^{2},
$$
$$
{\cal M}^{\parallel}({\bf k},{\bf k}_{1})
= \frac{g^2}{\vert {\bf k} \vert \, \vert {\bf k}_{1} \vert} \,
\frac{({\bf k}_2 {\bf v})}{{\bf k}_2^2}
\left( \frac{k_2^2 \Delta^l(k_2)}{\omega \omega_1 \omega^2_2}
\, \delta \Gamma^{ijl}(k,-k_1,-k_2)k^ik^j_1k^l_2 \right)_{\omega
= \omega_{\bf k}^{l}, \, \omega_1 = \omega_{{\bf k}_{1}}^{l}}.
\eqno{(9.9)}
$$
HTL-correction $\delta \Gamma_3$ enters in the amplitude (9.9)
only. Here the contribution of bare three-gluon vertex is dropped out,
since by the definition (7.5),
$\Gamma^{ijl}(k,-k_1,-k_2)k^ik^j_1k^l_2 \equiv 0.$

The quantity ${\cal M}^{\parallel}$ can be interpreted as
the scattering amplitude of a longitudinal
wave by dressing "cloud" of a particle.
However, in contrast to
the Abelian plasma here, the scattering is produced not by
the oscillation of a screening cloud of color charge
as a result of interaction with incident scattering wave,
but as a consequence the fact that it induced by precession of color vectors
of particles forming this cloud
in the incident wave field $\omega^l_{\bf k}$. This process of
transition scattering is a completely collective effect.
For calculation of its probability
${\it w}_{\bf v}^{\parallel}({\bf k}, {\bf k}_1)$ it is necessary to solve
the kinetic equation describing a color charges motion in a screening cloud in
the field that is equal to the sum of fields of incident wave (9.5)
and a "central" charge producing a screening cloud.

Now we consider the contribution with transverse virtual wave.
Using the association of projectors
$$
P_{\rho \alpha}(k_2) =
g_{\rho \alpha} - u_{\rho}u_{\alpha} - \frac{\omega_2^2}{k_2^2}
\tilde{Q}_{\rho \alpha}(k_2),
$$
we rewrite it in the form
$$
-2g^2N_{c} \int \, {\rm d} {{\bf k}_{1}}
\left( \frac{W_{{\bf k}_{1}}^{l}}{\omega^l_{{\bf k}_1}} \right)
\frac{1}{{\bf k}^{2}{\bf k}_{1}^{2}}
\Big( \frac{\partial {\rm Re} \, \varepsilon^{l}(k)}{\partial \omega}
\Big)_{\omega = \omega_{\bf k}^{l}}^{-1}
\Big( \frac{\partial {\rm Re} \, \varepsilon^{l}(k_{1})}{\partial \omega_1}
\Big)_{\omega_{1} = \omega_{{\bf k}_{1}}^{l}}^{-1}
$$
$$
\Big[ \frac{1}{( \omega \omega_1 )^2} {\rm Im} \,
( \Delta^t(k_2) \,^{\ast} \Gamma^{ijl}(k,-k_1,-k_2)k^ik^j_1
\,^{\ast} \Gamma^{i^{\prime}j^{\prime}l}(k,-k_1,-k_2)k^{i^{\prime}}
k^{j^{\prime}}_1 -
\eqno{(9.10)}
$$
$$
- \frac{1}{{\bf k}_2^2}
\Delta^t(k_2)( \,^{\ast} \Gamma^{ijl}(k,-k_1,-k_2)k^ik^j_1k^l_2)^2)
\Big]_{\omega = \omega_{\bf k}^{l}, \; \omega_{1} = \omega_{{\bf k}_{1}}^{l}}.
$$
To simplify the expression, we expand
$\,^{\ast} \Gamma^{ijl}k^ik^j_1$
in two mutually ortogonal vectors: ${\bf k}_2$ and $[{\bf n}{\bf k}_2]$,
where ${\bf n} \equiv [{\bf k}{\bf k}_1]$
\[
\,^{\ast} \Gamma^{ijl}k^ik^j_1=
(\,^{\ast} \Gamma^{ijs}k^ik^j_1k_2^s) \frac{k_2^l}{{\bf k}_2^2} +
(\,^{\ast} \Gamma^{ijs}k^ik^j_1[{\bf n}{\bf k}_2]^s)
\frac{[{\bf n}{\bf k}_2]^l}{{\bf n}^2 {\bf k}_2^2}.
\]
Substituting the last expresion into (9.10), instead of the expression
in square brackets, we define
$$
\Big[
\frac{1}{( \omega \omega_1)^2} \,
\frac{1}{{\bf n}^2 {\bf k}_2^2} \,
{\rm Im}
( \Delta^t(k_2)( \,^{\ast} \Gamma^{ijl}(k,-k_1,-k_2)k^ik^j_1
[{\bf n}{\bf k}_2]^l)^2)
\Big]_{\omega = \omega_{\bf k}^{l}, \, \omega_{1} = \omega_{{\bf k}_{1}}^{l}}.
$$
We transform the imaginary part similar to (9.8)
$$
{\rm Im}
( \Delta^t(k_2)( \,^{\ast} \Gamma^{ijl}k^ik^j_1
[{\bf n}{\bf k}_2]^l)^2) =
- {\rm Im} ( \Delta^{-1 \, t}(k_2))
\vert \Delta^t(k_2)\,^{\ast} \Gamma^{ijl}k^ik^j_1
[{\bf n}{\bf k}_2]^l \vert^2 +
$$
$$
+2( {\rm Im} \,^{\ast} \Gamma^{ijl}k^ik^j_1
[{\bf n}{\bf k}_2]^l)
{\rm Re}
( \Delta^t(k_2)\,^{\ast} \Gamma^{ijl}k^ik^j_1
[{\bf n}{\bf k}_2]^l).
\eqno{(9.11)}
$$
If we take into account ${\rm Im} \, \Delta^{-1 \, t}(k_2) = -(1/2) \, {\rm Im} \,
\Delta^{-1 \, l}(k_2)$, then the contribution of the first term in the r.h.s.
of (9.11)
to $\gamma^l({\bf k})$
can be presented in the form
$$
3( \omega_{pl}/g)^2 \int \, ( \omega_{\bf k}^{l} - \omega_{{\bf k}_
{1}}^{l}) {\it w}_{\bf v}^{\perp}({\bf k},{\bf k}_{1}) \Big( \frac{
W_{{\bf k}_{1}}^{l}}{\omega_{{\bf k}_1}^{l}} \Big)
\frac{{\rm d} \Omega}{4 \pi} \, {\rm d} {\bf k}_{1},
$$
where probability ${\it w}_{\bf v}^{\perp}({\bf k},{\bf k}_1)$ is obtained
from ${\it w}_{\bf v}^{\parallel}({\bf k},{\bf k}_1)$ with replacement
${\cal M}^{\parallel}({\bf k},{\bf k}_1)$ by
$$
{\cal M}^{\perp}({\bf k},{\bf k}_1) =
\frac{g^2}{ \vert {\bf k} \vert \vert {\bf k}_1 \vert}
\frac{([{\bf n}{\bf k}_2]{\bf v})}{{\bf n}^2{\bf k}_2^2}
\left( \frac{\Delta^t(k_2)}{\omega \omega_1}
\,^{\ast} \Gamma^{ijl}k^ik^j_1 [{\bf n}{\bf k}_2]^l \right)_{\omega=
\omega^l_{\bf k}, \; \omega_1= \omega^l_{{\bf k}_1}}.
\eqno{(9.12)}
$$
In contrast to preceding case, ${\cal M}^{\perp}$ represents the sum of
two contributions. The first one defined by bare three-gluon
vertex is connected with self-interaction of a gauge field and has no
analogy in Abelian plasma. The second contribution is associated
with the scattering of wave by the screening cloud of the charges.

The remaining terms in the r.h.s. of (9.8)
and (9.11) represent the interference of Thomson scattering
by longitudinal and transverse virtual waves, respectively. It is easily to see
this, having used
$$
{\rm Im}( \delta \Gamma^{ijl}k^ik^j_1k^l_2) =
-3 \pi \omega_{pl}^2 \, \omega_2^2
\int \, \frac{{\rm d} \Omega}{4 \pi} \, \frac{({\bf k}{\bf v})({\bf k}_1{\bf v})}
{vk} \, \delta(vk_2),
$$
$$
{\rm Im}( \,^{\ast} \Gamma^{ijl}k^ik^j_1[{\bf n}{\bf k}_2]^l) =
-3 \pi \omega_{pl}^2 \, \omega_2
\int \, \frac{{\rm d} \Omega}{4 \pi} \, \frac{({\bf k}{\bf v})({\bf k}_1{\bf v})}
{vk} ([{\bf n}{\bf k}_2]{\bf v})
\, \delta(vk_2).
$$
The interference of the last two mechanisms of scattering is absent.

Thus, summing the preceding, instead of (7.1), (7.2)
we have
$$
\gamma^{l}({\bf k}) = 3(\omega_{pl}/g)^2
\int \, ( \omega_{\bf k}^{l} - \omega_{{\bf k}_
{1}}^{l}) Q({\bf k},{\bf k}_{1}) \Big( \frac{
W_{{\bf k}_{1}}^{l}}{\omega_{{\bf k}_1}^{l}} \Big)
\, {\rm d} {\bf k}_{1},
\eqno{(9.13)}
$$
where
$$
Q({\bf k}, {\bf k}_{1}) =
2 \pi N_{c}
\bigg( \frac{\partial {\rm Re} \, \varepsilon^{l}(k)}{\partial \omega}
\bigg)_{\omega = \omega_{\bf k}^{l}}^{-1}
\bigg( \frac{\partial {\rm Re} \, \varepsilon^{l}(k_{1})}{\partial \omega_{1}}
\bigg)_{\omega_{1} = \omega_{\bf k_1}^{l}}^{-1}
\eqno{(9.14)}
$$
$$
\int \, \frac{{\rm d} \Omega}{4 \pi} \delta ( \omega_{\bf k}^{l}
- \omega_{{\bf k}_{1}}^l -
{\bf v}( {\bf k} - {\bf k}_{1}))
\, \vert {\cal M}^{c}({\bf k},{\bf k}_{1})
+ {\cal M}^{\parallel}({\bf k},{\bf k}_{1})
+ {\cal M}^{\perp}({\bf k},{\bf k}_{1}) \vert^{2}.
$$
The interference term in (9.14) between ${\cal M}^{\parallel}$ and
${\cal M}^{\perp}$ vanishes by relation
\[
k_2^i [{\bf n}{\bf k}_2]^j
\int \, \frac{{\rm d} \Omega}{4 \pi}v^iv^j \delta(vk_2) =0 .
\]

It is convenient to interpret the terms intering to
${\cal M} \equiv {\cal M}^c + {\cal M}^{\parallel} + {\cal M}^{\perp}$
using a quantum language. In this case the term ${\cal M}^c$
connected with the Thomson scattering can be represented as the
Compton scattering of the quantum of the soft modes (plasmon) by
QGP termal particle.
${\cal M}^{\parallel}$ defines the scattering of a quantum oscillation
through a longitudinal virtual wave with the propagator $\Delta^l(k_2)$,
where a vertex of a three-wave interaction is induced by HTL-correction
$\delta \Gamma_3$. ${\cal M}^{\perp}$ defines the quantum
oscillation scattering by transverse virtual wave with propagator
$\Delta^t(k_2)$. In this case ${\cal M}$ fulfils role of the total scattering
amplitude.

At the end of this Section we note the principal distinction of
obtained expression of the nonlinear Landau damping rate from the damping
rate for hard particles [31]. In the last case the scattering amplitude
involves the resummed gluon propagator in the electric and the magnetic
channels only, i.e. $\Delta^{l} (k_2)$ and $\Delta^{t} (k_2)$ respectively.
In our case, for small particle momenta $(\vert {\bf k} \vert \leq g T)$,
the scattering amplitude takes into account the thermal masses of the
particles and exchange contributions (vertex corrections).
This leads to more complicated expression for ${\cal M}$ involving
the effective vertecies, in contrast to similar
expression for the fast particles.
\vspace{1cm}\\
{\bf 10. THE ASSOCIATION WITH HTL-APPROXIMATION}\\

The kernel $Q({\bf k},{\bf k}_1)$ possesses two main properties.
The following inequality results from definition (9.14)
$$
Q({\bf k},{\bf k}_1) \geq 0.
\eqno{(10.1)}
$$
Next from (9.4) it follows that ${\cal M}^c({\bf k}, {\bf k}_1) =
{\cal M}^{\ast c}({\bf k}_1,{\bf k})$. The correctness of this equality
results from the conservation law (6.13). For ${\cal M}^{\parallel}$ and
${\cal M}^{\perp}$ we have the similar relations: ${\cal M}^{\parallel,
\perp}({\bf k},{\bf k}_1) =
({\cal M}^{\parallel, \perp}({\bf k}_1,{\bf k}))^{\ast}$.
Their proof trivially follows from the definitions of $\,^{\ast} \Gamma_3$
and $\Delta^{l,t}$. The consequence of these equalities is a main property
of a symmetry of kernel $Q({\bf k},{\bf k}_1)$ with respect to permutation of
a wave vectors ${\bf k}$ and ${\bf k}_1$
$$
Q({\bf k},{\bf k}_1) = Q({\bf k}_1,{\bf k}).
\eqno{(10.2)}
$$

From (9.13) follows that in the case of a global equilibrium plasma
and by inequality (10.1), waves of a high frequencies are damped out,
and a smaller ones are increased. In particular, in the limit of
$\vert {\bf k} \vert \rightarrow 0$ we obtain from (9.13)
$$
\gamma^l(0) = 3( \omega_{pl}/g)^2
\int \, ( \omega_{pl} - \omega^l_{{\bf k}_1})
Q(0,{\bf k}_1) \left( \frac{W^l_{{\bf k}_1}}{\omega^l_{{\bf k}_1}} \right)
{\rm d} {\bf k}_1 < 0,
$$
i.e. ${\bf k} =0$-mode is not damped, as it was calculated in Ref. [17].

This clearly shows that nonlinear Landau damping rate
(9.13) which is not of fixed sign, cannot be identified with gluon
damping rate, calculated on the basis of resumming Braaten-Pisarski
techniques. Actually $\gamma^l({\bf k})$ defines two processes:
the effective pumping of energy across the spectrum sideways of small
wave numbers with conservation of excitation energy and properly nonlinear
dissipation (damping) of longitudinal plasma waves by QGP particles,
where the first process is crucial. To see this, we rewrite the
kinetic equation (6.14) as follows
$$
\frac{\partial W^l_{\bf k}}{\partial t}
= - 3( \omega_{pl}/g)^2
\int \, ( \omega^l_{\bf k} - \omega^l_{{\bf k}_1}) \, \omega^l_{\bf k}
Q({\bf k},{\bf k}_1)
\left( \frac{W^l_{\bf k}}{\omega^l_{\bf k}} \right)
\left( \frac{W^l_{{\bf k}_1}}{\omega^l_{{\bf k}_1}} \right) {\rm d} {\bf k}_1.
\eqno{(10.3)}
$$
Hereafter, we have restricted ourseves to spatially-homogeneous case.

We perform replacement $\omega^l_{\bf k} =
( \omega^l_{\bf k} - \omega^l_{{\bf k}_1})/2 +
( \omega^l_{\bf k} + \omega^l_{{\bf k}_1})/2$,
then last equation can be presented as
$$
\frac{\partial W^l_{\bf k}}{\partial t}
= - \int \, Q_{S}({\bf k},{\bf k}_1)
W^l_{\bf k} W^l_{{\bf k}_1} {\rm d} {\bf k}_1 -
\int \, Q_{A}({\bf k},{\bf k}_1)
W^l_{\bf k} W^l_{{\bf k}_1} {\rm d} {\bf k}_1,
\eqno{(10.4)}
$$
where we introduce symmetric and antisymmetric kernel, respectively
$$
Q_{S}({\bf k},{\bf k}_1) \equiv 3( \omega_{pl}/g)^2
\frac{( \omega^l_{\bf k} - \omega^l_{{\bf k}_1})^2}
{2 \omega^l_{\bf k} \omega^l_{{\bf k}_1}}
Q({\bf k},{\bf k}_1), \;
$$
$$
Q_{A}({\bf k},{\bf k}_1) \equiv 3( \omega_{pl}/g)^2 \,
\frac{( \omega^l_{\bf k})^2 - ( \omega^l_{{\bf k}_1})^2}
{2 \omega^l_{\bf k} \omega^l_{{\bf k}_1}}
Q({\bf k},{\bf k}_1).
\eqno{(10.5)}
$$
We integrate Eq. (10.4) over ${\rm d} {\bf k}$ and introduce the
total energy of longitudinal QGP excitations: $W^l_0 = \int \, W^l_{\bf k}
{\rm d}{\bf k}.$ Then by the properties of kernels (10.5) we obtain
$$
\frac{\partial W^l_0}{\partial t}
= - \int \, Q_{S}({\bf k},{\bf k}_1)
W^l_{\bf k} W^l_{{\bf k}_1} {\rm d} {\bf k} {\rm d} {\bf k}_1 < 0,
$$
i.e. properly nonlinear dissipation of excitations is defined by symmetric
part of a kernel $Q_{S}$ only. It is necessary to compare
the parts of nonlinear Landau damping rate which correspond to the nonlinear
dissipation of waves only,
with damping rate
of boson modes from HTL-approximation, namely
$$
\gamma^l_{S}( {\bf k}) \equiv 3( \omega_{pl}/g)^2
\int \,
\frac{( \omega^l_{\bf k} - \omega^l_{{\bf k}_1})^2}
{2 \omega^l_{\bf k} \omega^l_{{\bf k}_1}}
\, Q({\bf k},{\bf k}_1)W^l_{{\bf k}_1} {\rm d} {\bf k}_1.
\eqno{(10.6)}
$$
The function $\gamma^l_{S}({\bf k})$ is positive for any value of a wave vector
${\bf k}$ and particular, for $\vert {\bf k} \vert =0$.

Antisymmetric part of a kernel is not associated with dissipative phenomenon
and defines the spectral pumping from short to long waves. It is easy to see,
by considering the model problem of interaction of two
infinitely narrow packets with typical wave vectors ${\bf k}_1, {\bf k}_2$.

Let us introduce $W^l_{\bf k}$ as follows
$$
W^l_{\bf k}(t) = W_1(t) \delta({\bf k} -{\bf k}_1) +
W_2(t) \delta({\bf k} -{\bf k}_2), \;
\vert {\bf k}_1 \vert > \vert {\bf k}_2 \vert.
$$
Substituting the last expression into (10.4), we obtan the coupled
nonlinear equations
$$
\frac{\partial W_1}{\partial t} = - (Q_S + Q_A) W_1W_2  \; \; , \; \;
W_1(t_0) = W_{10},
$$
$$
\frac{\partial W_2}{\partial t} = - (Q_S - Q_A) W_1W_2 \; \; , \; \;
W_2(t_0) = W_{20}.
$$
Here, $Q_{S,A} \equiv Q_{S,A}({\bf k}_1, {\bf k}_2)$. The general solution
of this system has the form
$$
W_1(t) = -W_{10} \, C
\frac{{\rm e}^{-C(t-t_0)}}
{-(Q_{S} + Q_{A})W_{20} + (Q_{S} - Q_{A})W_{10} \, {\rm e}^{-C(t-t_0)}} ,
$$
$$
W_2(t) = \frac{Q_{S} - Q_{A}}{Q_{S} + Q_{A}} \, W_1(t) -
\frac{C}{Q_{S} + Q_{A}}, \;
C \equiv (Q_{A} - Q_{S})W_{10} +
(Q_{A} + Q_{S})W_{20}.
$$
For $\vert{\bf k}_1 \vert > \vert{\bf k}_2 \vert$, by the definitions
(10.5), $C > 0$ and therefore, in the limit for
$t \rightarrow \infty $ we have
$$
W_1(t) \rightarrow 0, \;
W_2(t) \rightarrow W_0 - \frac{2Q_{S}}{Q_{S} + Q_{A}} W_{10}.
$$
Here $W_0 = W_{10} + W_{20}$ is total initial energy of packets. Thus we see
that as result of nonlinear interaction of two infinitely narrow packets the
effective pumping of energy across the spectrum sideways of small wave numbers
takes place.

In this case, as a result of the pumping, the part of excitations energy
(proportionaled to $Q_{S}$) nonlinear absorbs by QGP particles. The absorption
value converges to
$$
\Delta W = ( \omega^l_{{\bf k}_1} - \omega^{l}_{{\bf k}_2})
(W_{10}/ \omega^l_{{\bf k}_1}).
$$

The process of nonlinear scattering of plasmons by QGP particles only, not
results in their relaxation in homogeneous isotropic plasma. In fact, by the
kinetic equation (10.3) the general plasmons numbers conserves
$$
\frac{\partial N^l}{\partial t} =
\frac{\partial}{\partial t} \int \, N^l_{\bf k} {\rm d} {\bf k} =0.
$$
It follows that if for time $t=t_0$, $N^l_0$ plasmons are excited in a plasma,
then excitation energy for any $t \geq t_0$ does not less than a value
$\omega_{pl}N^l_0$ by conservation law of plasmon numbers. In homogeneous
plasma the total dissipation of longitudinal excitations energy is defind
by slow processes of four-wave interaction.
\vspace{1cm}\\
{\bf 11. THE ESTIMATION OF $\gamma^l_{S}(0)$}\\

Now we present a complete calculation of $\gamma^{l}_{S}({\bf k})$
at zero momentum of an incident field.
We start from representation of $\gamma^{l}_{S}({\bf k})$
in the form of (10.6) with kernel (9.14).
We introduce the coordinate system in which axis $0Z$
is aligned with vector ${\bf k}_1$; then the coordinates of vectors ${\bf k}$
and ${\bf {\bf v}}$ equal
${\bf k} = ( \vert {\bf k} \vert, \alpha, \beta),
{\bf v} = (1, \theta, \varphi),$ respectively.
By $\Phi$ we denote the angle between
${\bf v}$ and ${\bf k}$: $({\bf v}{\bf k}) = \vert {\bf k} \vert
\cos \Phi$. The angle $\Phi$ can be expressed as
$$
\cos \Phi = \sin \theta \sin \alpha \cos( \varphi - \beta) +
\cos \theta \cos \alpha.
\eqno{(11.1)}
$$

In the limit of $\vert {\bf k} \vert \rightarrow 0$, the kernel (9.14)
reduces to
$$
Q(0,{\bf k}_1) = \pi N_c \omega_{pl}
\frac{v^l_{{\bf k}_1} k_1^2}
{3 \omega_{pl}^2 - k_1^2}
\int \frac{{\rm d} \Omega}{4 \pi} \, \delta ( \cos \theta - \rho^l_{{\bf k}_1})
\vert {\cal M}^c(0,{\bf k}_1) + {\cal M}^{\parallel}(0,{\bf k}_1) +
{\cal M}^{\perp}(0,{\bf k}_1) \vert^2,
\eqno{(11.2)}
$$
where
\[
v^l_{{\bf k}_1} \equiv \omega^l_{{\bf k}_1}/ \vert {\bf k}_1 \vert, \;
k_1^2 \equiv (\omega^l_{{\bf k}_1})^2 - {\bf k}_1^2, \;
\rho^l_{{\bf k}_1} \equiv (\omega^l_{{\bf k}_1} - \omega_{pl})/
\vert {\bf k}_1 \vert \geq 0, \; {\rm d} \Omega= \sin \theta
{\rm d} \theta {\rm d} \varphi.
\]

By using the definitions (9.4), (9.9) and (9.12),
we find expressions
${\cal M}^{c}(0,{\bf k}_1)$, ${\cal M}^{\parallel}(0,{\bf k}_1)$ and
${\cal M}^{\perp}(0,{\bf k}_1)$. The first of them is defined more simply.
In the limit of zero momentum
$$
{\cal M}^c(0,{\bf k}_1) =
\left( \frac{g}{\omega_{pl}} \right)^2
\frac{\cos \Phi \cos \theta}{\omega^l_{{\bf k}_1}}.
$$

Calculation of ${\cal M}^{\parallel}(0,{\bf k}_1)$ is more complicated.
From (9.9) we obtain
$$
{\cal M}^{\parallel}(0,{\bf k}_1) =
\left( \frac{g}{\omega_{pl}} \right)^2
\frac{\omega_{pl}}{\omega^l_{{\bf k}_1}} \,
\frac{(1 - ( \rho^l_{{\bf k}_1})^2)}{{\bf k}^2_1  \rho^l_{{\bf k}_1}}
\lim \limits_{\vert {\bf k} \vert \to 0} \frac{1}{\vert {\bf k} \vert}
(\Delta^l(k_2) \delta \Gamma^{ijl}(k,-k_1,-k_2)k^ik^j_1k^l_2).
\eqno{(11.3)}
$$
Using the definition of HTL-correction (7.6) to bare three-gluon vertex,
after slightly cumbersome computations we define
$$
\lim \limits_{\vert {\bf k} \vert \to 0} \frac{1}{\vert {\bf k} \vert}
\delta \Gamma^{ijl}(k,-k_1,-k_2)k^{i}k_1^{j}k_2^{l} =
\eqno{(11.4)}
$$
$$
= - \cos \alpha \frac{{\bf k}_1^2}{\omega_{pl}} \,
\frac{( \rho^{l}_{{\bf k}_1})^3}{1-( \rho^{l}_{{\bf k}_1})^2}
\lim \limits_{\vert {\bf k} \vert \to 0}
\Delta^{-1 \, l}(k_2) + 3 \vert {\bf k}_{1} \vert^3 \cos \alpha \,
\rho^{l}_{{\bf k}_1}v^{l}_{{\bf k}_1}.
$$
In the last equility we use the definition of $\Delta^{-1 \, l}(k)$ as
the function $F( \omega / \vert {\bf k} \vert)$ (6.9).
Inserting (11.4) into (11.3),
we reduce the scattering amplitude with longitudinal virtual oscillation to
$$
{\cal M}^{\parallel}(0,{\bf k}_1) =
- \left( \frac{g}{\omega_{pl}} \right)^2
\frac{( \rho^l_{{\bf k}_1})^2}{\omega^l_{{\bf k}_1}} \, \cos \alpha
+ 3 \left( \frac{g}{\omega_{pl}} \right)^2
\omega_{pl} (1 - ( \rho^l_{{\bf k}_1})^2) \cos \alpha
\lim \limits_{\vert {\bf k} \vert \to 0}
\Delta^l(k_2).
\eqno{(11.5)}
$$

Now we consider the limit of the term with transverse virtual oscillation.
From (9.12) it follows
$$
{\cal M}^{\perp}(0,{\bf k}_1) =
\left( \frac{g}{\omega_{pl}} \right)^2
\frac{\omega_{pl}}{\omega^l_{{\bf k}_1}} \,
\frac{\cos \Phi - \cos \theta \cos \alpha}
{{\bf k}_1^2 \sin \alpha}
\lim \limits_{\vert {\bf k} \vert  \to 0}
\frac{1}{\vert {\bf k} \vert \vert {\bf n} \vert}
( \Delta^{t} (k_{2}) \,^{\ast} \Gamma^{ijl}
\, k^{i} k_1^{j} [ {\bf n} {\bf k_2} ]^{l}).
\eqno{(11.6)}
$$
Using the definition of effective three-gluon vertex (7.4)
and the relation
$$
(1 - ( \rho_{{\bf k}_1}^{l})^{2}) (1 - F( - \rho_{{\bf k}_1}^{l})) =
\frac{2}{3 \omega_{pl}^{2}}
\lim \limits_{\vert {\bf k} \vert \to 0}
\Delta^{-1 \, t} (k_{2}) + 1 +
\frac{2 {\bf k}_1^2}{3 \omega_{pl}^{2}}
(1 - ( \rho_{{\bf k}_{1}}^{l})^{2} ) \, ,
$$
we obtain
$$
\lim \limits_{\vert {\bf k} \vert \to 0}
\frac{1}{\vert {\bf k} \vert \vert {\bf n} \vert}
\, ^{\ast} \Gamma^{ijl} (k, - k_1, - k_2)
k^{i} k_{1}^{j} [ {\bf n} {\bf k}_{2} ]^{l} =
$$
$$
= - \frac{3}{2} \, \frac{{\bf k}_{1}^{4}}{\omega_{pl}}
\sin \alpha \, (1 - v_{{\bf k}_{1}}^{l} \rho_{{\bf k}_{1}}^{l}) -
\frac{{\bf k}_{1}^{2} \rho_{{\bf k}_{1}}^{l}}{\omega_{pl}}
\lim  \limits_{\vert {\bf k} \vert \to 0}
\Delta^{-1 t} (k_{2} ) \sin \alpha .
\eqno{(11.7)}
$$

Substitute (11.7) into (11.6) and take into account the
relation (11.1), we obtained required limit
$$
{\cal M}^{\perp}(0,{\bf k}_1) =
- \left( \frac{g}{\omega_{pl}} \right)^2
\frac{\rho^l_{{\bf k}_1}}{\omega_{{\bf k}_1}}
\sin \theta \sin \alpha \cos ( \varphi - \beta) -
\eqno{(11.8)}
$$
$$
- \frac{3}{2} \left( \frac{g}{\omega_{pl}} \right)^2 \vert {\bf k}_1 \vert
((v^l_{{\bf k}_1})^{-1} - \rho^l_{{\bf k}_1})
\sin \theta \sin \alpha \cos ( \varphi - \beta)
\lim \limits_{\vert {\bf k} \vert \to 0}
\Delta^{t}(k_2).
$$

The terms in the amplitude ${\cal M}$ not containing $\Delta^{l,t}$
combine to give
$$
\left( \frac{g}{\omega_{pl}} \right)^2
\frac{1}{\omega^l_{{\bf k}_1}}
\{ \cos \Phi \cos \theta - ( \rho^l_{{\bf k}_1})^2 \cos \alpha -
\rho^l_{{\bf k}_1} \sin \theta \sin \alpha \cos ( \varphi - \beta) \}.
$$
By the $\delta$-function in (11.2), the expression in curly braces
vanishes. Thus, all terms in ${\cal M}$, not containing the factors
$\Delta^{l,t}$, are in relatively reduced in the limit of ${\bf k}=0$-mode.

The remaining terms, after substitution into (11.2) and integration
over solid angle, yield
$$
Q(0,{\bf k}_1) = \frac{9}{2} \pi g^4 \frac{N_c}{ \omega_{pl}}
\, \frac{k_1^2 (1 - ( \rho_{{\bf k}_1}^l )^2 )}
{(3 \omega_{pl}^2 - k_1^2)} \, v_{{\bf k}_1}^{l} \theta(1 -
\rho_{{\bf k}_1}^{l} )
\{ (1 - (\rho_{{\bf k}_1}^{l})^2) \; \cos^2 \alpha
\vert \lim \limits_{\vert {\bf k} \vert \to 0} ( \Delta^{l}(k_2)) \vert^2 +
$$
$$
+ \frac{{\bf k}_1^2}{8 \omega_{pl}^2}
( (v_{{\bf k}_1}^l)^{- 1} - \rho_{{\bf k}_1}^l )^2
\; \sin^2 \alpha
\vert \lim \limits_{\vert {\bf k} \vert \to 0}
( \Delta^{t} (k_2)) \vert^{2} \} .
\eqno{(11.9)}
$$
We note that this expression is not dependent on angle $\beta$. This enables
us to represent the integration measure in the r.h.s. of (10.6)
in the form
$$
\int \, {\rm d} {\bf k}_{1} = 2 \pi \int\limits_{0}^{\infty} {\bf k}_1^2
{\rm d} \vert {\bf k}_{1} \vert
\int\limits_{1}^{- 1} \, {\rm d} ( \cos \alpha ).
$$

Futher we suppose that the excitations in QGP become
isotropic over the directions of vector ${\bf k}_1$ on a time scale
which is much less than time scale
of the nonlinear interaction. This enables us to consider the spectral
density  $W_{{\bf k}_1}^l$ as a function of $\vert {\bf k}_1 \vert$
only. Now we introduce the spectral function
$$
W^l_{\vert {\bf k}_1 \vert} \equiv
4 \pi {\bf k}_1^2
W^l_{{\bf k}_1}
$$
such that the integral $\int_{0}^{\infty} W^l_{\vert {\bf k}_1 \vert}
{\rm d} \vert {\bf k}_1 \vert = W^l$
is total energy of longitudinal oscillations in QGP. Substituting
(11.9) into (10.6) (for ${\bf k}=0$)
and performing the angular integration over $\alpha$, we obtain finally
$\gamma^l_{S}(0)$
$$
\gamma^{l}_{S}(0) \cong \int\limits_{0}^{\vert {\bf k}_1^{\ast} \vert}
Q ( \vert {\bf k}_1 \vert )
W_{\vert {\bf k}_1 \vert}^{l} \, {\rm d} \vert {\bf k}_1 \vert ,
\eqno{(11.10)}
$$
where a kernel $Q( \vert {\bf k}_1 \vert )$ has the form
$$
Q( \vert {\bf k}_1 \vert ) = \frac{9}{4} \pi g^2 N_c
\, \frac{k_1^2 (1 - ( \rho_{{\bf k}_1}^l )^2 )}
{(3 \omega_{pl}^2 - k_1^2)} \, \vert {\bf k}_1 \vert
( \rho_{{\bf k}_1}^{l})^2 \, \theta(1 -
\rho_{{\bf k}_1}^{l} )
\{ (1 - (\rho_{{\bf k}_1}^{l})^2)
\; \vert \lim \limits_{\vert {\bf k} \vert \to 0} ( \Delta^{l}(k_2)) \vert^2 +
$$
$$
+ \frac{{\bf k}_1^2}{4 \omega_{pl}^2}
( (v_{{\bf k}_1}^l)^{- 1} - \rho_{{\bf k}_1}^l )^2
\vert \lim \limits_{\vert {\bf k} \vert \to 0}
( \Delta^{t} (k_2)) \vert^{2} \} ,
\eqno{(11.11)}
$$
and the upper cutoff $\vert {\bf k}_1^{\ast} \vert$
distinguishes between soft and hard momenta:
$g T \ll \vert {\bf k}_1^{\ast} \vert \ll T$.
For crude estimation of $\gamma_S^l(0)$
let us expand the kernel $Q( \vert {\bf k}_1 \vert )$
in the momentum $\vert {\bf k}_1 \vert$,
using the approximations
$$
\omega_{{\bf k}_1}^l \approx \omega_{pl} , \, \,
\rho_{{\bf k}_1}^l \approx \frac{3 \vert {\bf k}_1 \vert}{10 \omega_{pl}} ,
\, \,
v_{{\bf k}_1}^l \approx \frac{\omega_{pl}}{\vert {\bf k}_1 \vert} ,
\, \, \, \ldots \; \; .
$$
Keeping the leading in $\vert {\bf k}_1 \vert$ term in the expansion
(11.11), we obtain
$$
Q ( \vert {\bf k}_1 \vert ) \approx  \frac{9 \pi}{800} \, N_c g^2 \,
\left( \frac{\vert {\bf k}_1 \vert}{\omega_{pl}^2} \right)^3 .
\eqno{(11.12)}
$$
The function $W_{{\bf k}_1}^l$ is approximated by its equilibrium value
[17]:
$W_{{\bf k}_1}^l \approx 4 \pi T$ and therefore
$$
W_{\vert {\bf k}_1 \vert}^l \approx 16 \pi^2 {\bf k}_1^2 \, T .
\eqno{(11.13)}
$$
Substituting (11.12) and (11.13) into
(11.10), and setting $\vert {\bf k}_1^{\ast} \vert \sim g T$,
we finally define
$$
\gamma^l_{S} (0) \approx + 1.04 N_c g^2 T .
\eqno{(11.14)}
$$
In our case the coefficient of the $g^2 T$ has the same sign but is
significantly large than
corresponding one, calculated in [4].\\

{\bf 12. THE $\gamma^{l}(0)$ DEPENDENCE ON A GAUGE
PARAMETER}\\

Now we estimate the contribution of a gauge dependence term of a
propagator in covariant gauge (6.8) to the nonlinear Landau damping rate
$$
{\cal D}^{\xi}_{\rho \alpha}(k_2) =
\xi k_{2 \rho} k_{2 \alpha} /[( \omega_2 + i \epsilon)^2 -
{\bf k}^2_2]^2, \; k_2 =k - k_1.
\eqno{(12.1)}
$$
By the Ward identities (7.10), a gauge-dependent part of
$\gamma^l({\bf k})$ can be represented as
$$
\gamma^l_{\xi}({\bf k}) =
-2 \xi g^2 N_c {\rm Im} \int \, {\rm d}{\bf k}_1 \bigg( \frac{
W^l_{{\bf k}_1}}{\omega^l_{{\bf k}_1}} \bigg) \Big[
\bigg( \frac{\partial {\rm Re} \, \varepsilon^l(k)}{\partial \omega} \bigg)^{-1}
\bigg( \frac{\partial {\rm Re} \, \varepsilon^l(k_1)}{\partial \omega_1} \bigg)^{-1}
$$
$$
\frac{( \omega \omega_1 ({\bf k}{\bf k}_1) - {\bf k}^2{\bf k_1}^2)^2}
{{\bf k}^2{\bf k}^2_1 (k^2k^2_1)^2}
( \Delta^{-1 \,l}(k) - \Delta^{-1 \,l}(k_1))^2
\frac{1}{[( \omega_2 + i \epsilon)^2 - {\bf k}^2_2]^2}
\Big]_{\omega = \omega^l_{\bf k}, \; \omega_1 = \omega^l_{{\bf k}_1}}.
\eqno{(12.2)}
$$
If we take into account that
$\Delta^{-1 \,l}(k) \vert_{\omega = \omega^l_{\bf k}} =
\Delta^{-1 \,l}(k_1) \vert_{\omega_1 = \omega^l_{{\bf k}_1}} \equiv 0$,
then formally, a gauge-dependent part of nonlinear Landau damping rate
vanishes, as it was mentioned in Sec. 8. However, we show that the integral
in the r.h.s. of (12.2) develops on mass-shell poles.
We consider, for example, the coefficient of $( \Delta^{-1 \, l}(k))^2$.
In the limit $\vert {\bf k} \vert \rightarrow 0$ this coefficient is equal to
$$
\int \, {\bf k}_1^2 {\rm d} \vert {\bf k}_1 \vert \bigg( \frac{
W^l_{{\bf k}_1}}{\omega^l_{{\bf k}_1}} \bigg)
\bigg( \frac{\omega^l_{{\bf k}_1}}{k^2_1} \bigg)^2
\bigg( \frac{\partial {\rm Re} \, \varepsilon^l(k_1)}{\partial \omega_1}
\bigg)^{-1}_{\omega_1 = \omega^l_{{\bf k}_1}}
{\rm Im}
\frac{1}{[( \omega^l_{{\bf k}_1} - \omega_{pl} - i \epsilon)^2 - {\bf k}^2_1]^2}
$$
(numerical factor is omitted).
For lower limit, integrand expression
(for $W^l_{{\bf k}_1} \simeq {\rm const}$) is as follows
$$
\int \limits_{\vert {\bf k}_1 \vert \simeq 0}
\frac{{\rm d} \vert {\bf k}_1 \vert}{\vert {\bf k}_1 \vert^2} ,
$$
i.e. the integral involves a power infrared divergence and is infinite at
the pole. Thus, a gauge parameter in (12.2) is multiplied by
$0 \times \infty$ uncertainty.  We investigate this uncertainty, following
by reasoning [27].

In order to evalute the double poles in (12.1) we use the
prescription
$$
\frac{1}{[(\omega_2 + i \epsilon)^2 - {\bf k}_2^2]^2} =
\lim\limits_{m^2\to 0} \frac{\partial}{\partial m^2}
\frac{1}{[(\omega_2 + i \epsilon)^2 - {\bf k}_2^2 - m^2]} =
$$
$$
= \lim\limits_{m^2\to 0} \frac{\partial}{\partial m^2} \,
\frac{1}{2 \sqrt{{\bf k}^2_2 + m^2}}
\bigg ( \frac{1}{\omega_2 - \sqrt{{\bf k}^2_2 + m^2} + i \epsilon} -
 \frac{1}{\omega_2 + \sqrt{{\bf k}^2_2 + m^2} + i \epsilon} \bigg ).
\eqno{(12.3)}
$$

By computing we first take $m^2 \rightarrow 0$ befor the effective on mass-shell
limit $\omega \rightarrow \omega_{pl}$. We focus on the damping rate
$\gamma_{\xi}^l$ of excitation at rest (at vanishing three-momentum).

Performing the interesting limit $\vert {\bf k} \vert \rightarrow 0$
in (12.2)
and taking
into account (12.3), we define
$$
\gamma_{\xi}^l (0) =
\frac{(2 \pi)^2}{3} \xi g^2 N_c \frac{1}{\omega_{pl}}
\lim \limits_{m^2 \to 0} \frac{\partial}{\partial m^2}
\int \limits_0^\infty
\frac{{\bf k}_1^2 {\rm d} \vert {\bf k}_1 \vert}
{\sqrt{{\bf k}_1^2 + m^2}}
\bigg ( \frac{W_{{\bf k}_1}^l}{\omega_{{\bf k}_1}^l} \bigg )
\bigg ( \frac{\partial {\rm Re} \,
\varepsilon^l(k_1)}{\partial \omega_1} \bigg)^{-1}_{\omega_1
= \omega_{{\bf k}_1}^l}
\bigg( \frac{\omega_{{\bf k}_1}^l}{k_1^2} \bigg)^2
$$
$$
( \Delta^{-1 \, l}(\omega, 0) - \Delta^{-1 \, l}
( \omega_{{\bf k}_1}^l, {\bf k}_1) )^2
\{ \delta ( \omega - \omega_{{\bf k}_1}^l - \sqrt{{\bf k}^2_1 + m^2} ) -
\delta ( \omega - \omega_{{\bf k}_1}^l + \sqrt{{\bf k}_1^2 +m^2}) \}.
\eqno{(12.4)}
$$
Without restriction for the general case we choose
$\omega > \omega_{pl} +m$. Then we obtain from
(12.4)
$$
\gamma_{\xi}^l (0) =
\frac{2 \pi^2}{3} \xi g^2 N_c \frac{1}{\omega_{pl}}
\lim \limits_{m^2 \to 0} \frac{\partial}{\partial m^2}
\bigg[ \bigg( \frac{k_{10}}{\sqrt{k_{10}^2 + m^2}} \bigg)
\bigg( \frac{W_{k_{10}}}{\omega_{k_{10}}} \bigg)
\bigg ( \frac{\partial {\rm Re} \, \varepsilon^l (k_{10})}
{\partial \omega_1}
\bigg)^{-1}_{\omega_1 = \omega_{{\bf k}_{10}}
^l}
\frac{(\omega_{k_{10}}^l)^2}{[( \omega_{k_{10}})^2 - k_{10}^2 ]^2}
$$
$$
( \Delta^{-1 \, l}(\omega, 0) -
\Delta^{-1 \, l}( \omega - \sqrt{k_{10}^2 + m^2}, k_{10}))^2
\bigg(
\frac{ \partial \sqrt{{\bf k}_1^2 + m^2}}{\partial {\bf k}_1^2} +
\frac{\partial \omega_{{\bf k}_1}^l}{\partial {\bf k}_1^2} \bigg)^{-1}_{
\vert {\bf k}_1 \vert = k_{10}} \bigg ] ,
\eqno{(12.5)}
$$
where $k_{10}$ is a solution of equation
$$
\omega_{{\bf k}_1}^l = \omega - \sqrt{{\bf k}_1^2 + m^2} .
$$
In order to perform the interesting limit $m^2 \rightarrow 0$ and
$\omega \rightarrow \omega_{pl}$, we notice that the solution $k_{10}$
vanishes for $m^2 = 0$ as
$$
k_{10} \simeq \omega - \omega_{pl} + O(k_{10}^2).
$$
Working out the derivative with respect to $\partial / \partial m^2$
and taking $m^2 \rightarrow 0$, we find that the most singular term in the
limit $k_{10} \rightarrow 0$ comes from the derivative
$$
\frac{\partial}{\partial m^2} \bigg(
\frac{k_{10}}{\sqrt{k^2_{10} + m^2}} \bigg) \rightarrow
- \frac{1}{2 k_{10}^2} .
$$
Using the approximation
$W_{{\bf k}_{10}}^l \approx 4 \pi T$ and
$\Delta^{- 1 \, l} (\omega, 0) - \Delta^{- 1 \, l}
(\omega - k_{10}, k_{10}) \approx
2 \omega _{pl} k_{10}$, we finally obtain
$$
\gamma_{\xi}^{l} (0) \simeq - 2 \frac{(2 \pi)^3}{3} \xi g^2 N_c T
\bigg ( \frac{\omega - \omega_{pl}}{\omega_{pl}} \bigg ).
\eqno{(12.6)}
$$
From the last expression we notice that by going on mass-shell, the
gauge-dependent part of the nonlinear Landau damping rate vanishes.

Excess factor $(\omega - \omega_{pl})$ of numerator (12.6)
is arised from the function
$$
\bigg(
\frac{\partial \sqrt{{\bf k}_1^2 + m^2}}{\partial {\bf k}_{1}^{2}} +
\frac{\partial \omega_{{\bf k}_1}^l}{\partial {\bf k}_{1}^{2}} \bigg)
_{\vert {\bf k}_1 \vert = k_{10}}^{- 1}
$$
in the expression (12.5). In paper [27]
this factor is compensated by singularity of statistic factor
$$
\lim \limits_{k_{10} \rightarrow 0}
\lim \limits_{m^2 \rightarrow 0} f_g ( \sqrt{k_{10}^2 + m^2} ) \simeq
\frac{T}{( \omega - \omega_{pl} )} ,
$$
associated with the spectral representation of bare propagator
$1/(k^2 + m^2)$. In our case this factor is absent.

One can lead to the result (12.6) in a different way.
In derivation of the expression for $\gamma^l({\bf k})$ (6.15)
we use representation of the spectral density $I_{\omega, {\bf k}}^l$
in the form (6.10). General speaking, this representation holds
in the linear approximation only, when the time correlation
(dependence on $\omega$) is one-to-one correspondence with the spatial
correlation (dependence on ${\bf k}$). To include the effects of weakly
nonstationary
ingomogeneous plasma motions (when one-to-one correspondence between
excitations frequency and its wave number fails), we replace the sharp
$\delta$-function in the $I_{\omega, {\bf k}}^{l}$ by Breit-Wigner form
[6, \,32, \,29], with width $\gamma^{l}$
$$
I^l_{\omega, {\bf k}} = \frac{1}{\pi}
\Big( I^l_{\bf k} \frac{\gamma^l}{( \omega - \omega^l_{\bf k})^2
+ ( \gamma^l)^2} +
I^l_{-{\bf k}} \frac{\gamma^l}{( \omega + \omega^l_{\bf k})^2
+ ( \gamma^l)^2} \Big).
\eqno{(12.7)}
$$
The parameter $\gamma^l$
is interpreted as damping rate of boson mode of order $g^2T$.
Note that in this case we are dealing with correlator
$\langle A^a_{\mu}(X_1)A^b_{\nu}(X_2) \rangle$ dependence on difference $X_1 - X_2$.
Dependence on the midpoint $(X_1 + X_2)/2$ is accounted in the form of a
dependence on a slow coordinate and a slow time in the l.h.s. of
kinetic equation (6.14) by means of the operator
${\partial}/{\partial t} + {\bf V}^l_{\bf k}{\partial}/{\partial {\bf x}}$.

Using representation (12.7) and formula
$$
\frac{1}{[( \omega_1 - \omega_{pl} - i \epsilon)^2 -{\bf k}^2_1]^2} =
\frac{\partial}{\partial {\bf k}_1^2} \,
\frac{1}{2 \vert {\bf k}_1 \vert}
\left( \frac{1}{\omega_1 - \omega_{pl} - \vert {\bf k}_1 \vert - i \epsilon} -
\frac{1}{\omega_1 - \omega_{pl} + \vert {\bf k}_1 \vert - i \epsilon} \right),
$$
in this case we derive, instead of (12.5)
$$
\gamma_{\xi}^l (0) =
- \frac{\pi}{3} \xi g^2 N_c \frac{1}{\omega^3_{pl}}
\int\limits_0^\infty
{\bf k}_1^2  {\rm d} \vert {\bf k}_1 \vert \,
W^l_{{\bf k}_1}
\frac{\partial}{\partial {\bf k}_1^2}
\Big[ \frac{1}{\vert {\bf k}_1 \vert}
\int \limits_{- \infty}^{+ \infty} \, {\rm d} \omega_1
\frac{{\gamma}^l}{( \omega_1 - \omega_{pl})^2 + ( \gamma^l)^2}
$$
$$
( \Delta^{-1 \, l}(\omega, 0) - \Delta^{-1 \, l}
(\omega_1, 0) )^2
\{
\delta (\omega_1 - \omega_{pl} - \vert {\bf k}_1 \vert) -
\delta (\omega_1 - \omega_{pl} + \vert {\bf k}_1 \vert) \}.
$$
Perform the integration with respect to over ${\rm d} \omega_1$,
and differentiation with respect to
$\vert {\bf k}_1 \vert^2$, we obtain expression
$$
\gamma_{\xi}^{l} (0) \simeq - 4^2 \frac{(2 \pi)^2}{3} \xi g^2 N_c T
\bigg ( \frac{\omega - \omega_{pl}}{\omega_{pl}} \bigg )
\int \limits^{\infty}_0 \,
{\bf k}_1^2 {\rm d} \vert {\bf k}_1 \vert
\frac{\gamma^l}{[{\bf k}^2_1 + ( \gamma^l)^2 ]^2}.
$$
It follows the result (12.6).
\vspace{1cm}\\
{\bf 13. CONCLUSION}\\

Let us consider in more detail approximations scheme, which we use in this
paper. In fact, here two types of the approximations are used. The first
of them is connected with the employment of usual approach, developed
in Abelian plasma, to QGP, i.e. the standard expansion of the current in powers of
the oscillations amplitude and computation of interacting field in the form of a
series of a perturbation theory in powers of a free field $A^{(0)}$
(more precisely, in
$g A^{(0)}$). However, in contrast to Abelian plasma,
in our case even if the first two nonlinear orders of the color current
are taken into account, much more terms, defining the nonlinear scattering of
waves are derived. Here, we use second approximation,
connected with the notions going from the papers devoted to high-temperature
QCD or more precisely, the set of rules for power counting these
terms, developed
by Blaizot and Iancu [12]. These rules enable us to singled out
the leading terms in the coupling constant.
These terms are purely non-Abelian in a full accordance with the
conclusions of Ref. [12].

However we note that in [12] behaviour of a mean field $\langle
A_{\mu}^{a}(X) \rangle$ is investigated, which in our case vanishes.
In the covariant derivative ${\cal D}_{\mu}$, containing only the
random part of a gauge field in our approach, we suppose that
$\partial / \partial X^{\mu}$ and
$gA_{\mu}$ are of the different orders. In field strength tensor
we distinguish the linear and nonlinear parts. Therefore the
aproximatons scheme which have been carried out in this paper suffers
from disadvantage of breaking
non-Abelian gauge symmetry of a theory at each step of
approximate calculation, as it was discussed in Sec. 4.

This is really true if we suppose that the magnitude of a soft
random oscillations $\vert A_{\mu}^a(X) \vert$ is of order $T$ or
$\vert A_{\mu}^a (k) \vert \sim 1/g (gT)^3$ in the Fourier representation.
In this case, by using obtained expressions for terms in the expansion of a
colour current (3.2), (5.4),
(5.9) and estimations (5.7), (5.12), we have
$$
j_{\mu}^{T (1)} (k) \sim
j_{\mu}^{T (2)} (k) \sim
j_{\mu}^{T (3)} (k) \sim \; \ldots \; \sim \frac{1}{g^2 T},
$$
i.e. all terms in the expansion (2.14) are of the same order in
magnitude and the problem of resummation of all the relevant contributions
appears. Thus a gauge symmetry is recovered.

In this paper we have restricted ourselves to just a
finite number of terms in the expansion (2.14).
This impose more rigorous restriction on the magnitude of random oscillations:
$\vert A_{\mu}^a (X) \vert \sim gT \; \; ( \vert A_{\mu}^a (k) \vert
\sim 1/(gT)^3)$. In this case we have
$$
j^{T(1)}_{\mu} (k) \sim \frac{1}{gT} \; , \;
j^{T(2)}_{\mu} (k) \sim \frac{1}{T} \; , \;
j^{T(3)}_{\mu} (k) \sim \frac{g}{T} \; , \ldots,
$$
each following term in the random current expansion is suppressed
by more powers of $g$ and use of the perturbation theory is justified.
A gauge symmetry is restored when we take into account consistently
all contributions at the leading order in $g$
(Sec. 6) to the probability of the soft excitations
scattering by the hard thermal particles.
In this case we derive
the expression for the nonlinear Landau damping rate (7.1), (7.2)
which is closely allied in the form to corresponding gluon damping rate
in HTL-approximation [4].

In Sec. 10 it was shown that the nonlinear interaction of longitudinal
eigenwaves leads to effective pumping of energy across the spectrum sideways
of small wave numbers.
Consequence of this fact is the inequality $\gamma^l(0) < 0$, i.e.
 ${\bf k}=0$ -
mode is increased. The dissipation of the energy of plasma waves
by QGP particles here, is not lead to total relaxation of a plasma excitations.
In the scale of a small $\vert {\bf k} \vert$, effects
described by nonlinear terms in the expansion of the colour current of
higher-order in the field, come into play. Consideration these effects
involves suppression of increase of ${\bf k}=0$-mode.

The r.h.s. of obtained kinetic equation (6.16)
in the regime
$\vert {\bf k} \vert \ll g T$ contains
the part of possible processes of a plasmon scattering in QGP only,
namely, the processes of a type
$$
g^{\ast} + g \rightarrow   g^{\ast}  + g \; ,
$$
$$
g^{\ast} + q(\bar{q})  \rightarrow g^{\ast} + q (\bar{q}) \; ,
$$
where $g^{\ast}$ is plasmon collective excitations and $g, q, \bar{q}$
are excitations with characteristic momenta of order $T$.
Diagrammatically this corresponds to graph with four external lines,
where one of the incoming (outcoming) lines is a soft and the other is a hard.
However it is clear that there are further contributions to the damping
rate of a soft gluon, going without exchange of energy between a hard particles
and waves. They are associated with the processes of nonlinear plasmon
scattering by a soft excitations of QGP, i.e. with the processes of a type
$$
g^{\ast} + g^{\ast} \rightarrow g^{\ast} + g^{\ast} \; ,
\eqno{(13.1)}
$$
$$
g^{\ast} + q^{\ast} ( \bar{q}^{\ast} ) \rightarrow
g^{\ast} + q^{\ast} ( \bar{q}^{\ast} ) \; ,
\eqno{(13.2)}
$$
where $q^{\ast}, \bar{q}^{\ast}$ are plasmino collective excitations.
The kinetic equation describing the process (13.1) and
(13.2) is the equation of a purely Boltzmann type, i.e.
collision term in the r.h.s. of this equation has standard
Boltzmann structure, with on gain term and a loss term. The probability
of these processes is defined by preceding methods from nonlinear current
of a fourth order $j_{\mu}^{T (4)}$, if the process of interaction iteration
of higher-order in the field is taken into account.

As it was mentioned in Introduction, the Boltzmann equation was already used for
definition of damping rate of the fast particles. In paper [11] (see also [14])
on the basis of Boltzmann equation, the damping rate for hard gluons in
the leading logarithmic order has been computed. The value of obtained
the damping rate is fully coincident with corresponding damping rate derived in
quantum theory [6, \,7].
The scattering matrix element appearing in collision term corresponds to
elastic scattering of hard gluons in the resummed Born approximation.
By Heiselberg and Pethick [11]
was noted that the particle damping rate is considerably more difficult
to calculate for small particle momenta, $\vert {\bf k} \vert \leq g T$.
Particles momentum becomes of the same order with momentum transfers.
As in the case of scatterng of plasmons by hard particles discussed above,
vertex corrections should be taken into account also. It leads to more
complicated expressions for probabilities of both plasmon-plasmon (13.1)
and plasmon-plasmino (13.2) scattering in contrast to (9.14).

A detailed research of the process of (13.1) and its influence to
relaxation of soft Bose excitations in QGP will be presented somewhere [26].
Here, we note only, that it is the process of higher-order in the field,
which probability is defined with the help of three-gluon, four-gluon
effective vertecies and effective propagator, as the probability of the
process of nonlinear scattering of a plasmon by QGP particles, derived
in this paper.
\vspace{1cm}\\
{\bf ACKNOWLEDGEMENTS}\\

This work was supported by the Russian Foundation for Basic Research
(Project No.97-02-16065).
\vspace{3cm}\\
{\bf APPENDIX A}\\

We rewrite the system of equations (9.6), (9.7)
in the coordinate represetation. In the temporal gauge we have
$$
\frac{{\rm d}^2 {\bf x}}{{\rm d}t^2} =
\frac{g}{{\rm m}} \sqrt{1 - {\bf v}^2} \, {\rm Q}^a
({\bf E}^a - {\bf v}({\bf v}{\bf E}^a)),
$$
$$
\frac{{\rm d}{\rm Q}^a}{{\rm d}t} =
gf^{abc}({\bf v}{\bf A}^b) \, {\rm Q}^c,
\eqno{(A1)}
$$
where $t$ is a coordinate time; ${\bf v}={\rm d}{\bf x}/{\rm d}t, \,
{\bf E}^a = - \partial {\bf A}^a/ \partial t$, and
$$
{\bf A}^a({\bf x},t) =
\int {\bf A}^a_{\bf k} \, {\rm e}^{i{\bf k}{\bf x}
- i \omega^l_{\bf k}t} {\rm d}{\bf k} \; ,
\; {\bf A}^a_{\bf k} = \frac{{\bf k}}{\vert {\bf k} \vert} A^a_{\bf k}.
\eqno{(A2)}
$$
The Eq. (9.6) corresponding to the component $\mu = 0$,
becomes identity.

If we neglect by the wave action on typical particle, then the colour charge
motion will be constant
and its colored vector will be fixed with initial condition:
${\bf x}_0 (t) = {\bf v}_0t, \, {\rm Q}^a = {\rm Q}^a_0.$
To derive the oscillations of a color particle, excited by fields and
being linear order in amplitude of field, it is necessary to neglect
by variations of ${\bf v}$, ${\rm Q}^a$
in the r.h.s. of $(A1)$ and set
$$
{\bf A}^a ({\bf x}, t) \simeq {\bf A}^a ({\bf v}_0 t, t) =
\int {\bf A}^a_{\bf k} \, {\rm e}^{- i(\omega^l_{\bf k} - {\bf v}_0 {\bf k})t}
\, {\rm d} {\bf k} ,
$$
instead of $(A2)$.

In this approximation the solution of system $(A1)$ has the form
$$
{\bf x} (t) = {\bf v}_0 t + \left( - \frac{g}{{\rm m}} \right) {\rm Q}_0^a
\sqrt{1 - {\bf v}_0^2} \int \,
\frac{{\bf E}_{\bf k}^a - {\bf v}_0 ({\bf v}_0 {\bf E}_{\bf k}^a)}
{(\omega_{\bf k}^l - {\bf v}_0 {\bf k})^2}
\, {\rm e}^{-i (\omega_{\bf k}^l - {\bf v}_0 {\bf k}) t} \, {\rm d} {\bf k}
\equiv {\bf x}_0 (t) + \Delta {\bf x} (t) ,
$$
$$
{\rm Q}^a(t) = {\rm Q}_0^a + i g f^{abc} {\rm Q}_0^c
\int \,
\frac{({\bf v}_0 {\bf A}_{\bf k}^b)}
{\omega_{\bf k}^l - {\bf v}_0 {\bf k}}
\, {\rm e}^{-i (\omega_{\bf k}^l - {\bf v}_0 {\bf k}) t} \, {\rm d} {\bf k}
\equiv {\rm Q}_0^a + \Delta {\rm Q}^a (t).
\eqno{(A3)}
$$
Using these expressions we derive the radiation intensity of
oscillating colour charge.

It is equal to work of radiation field with charge in unit time
$$
{\cal W}^l = \int \, ({\bf E}_{{\bf Q}}^a \,
{\bf j}_{{\bf Q}}^a) \, {\rm d} {\bf x} ,
\eqno{(A.4)}
$$
where ${\bf E}_{{\bf Q}}^a$ is a field induced by a colour current
${\bf j}_{{\bf Q}}^a$ of a charge ${\rm Q}^a$. The sign in the r.h.s.
of $(A4)$ corresponds to choose of a sign in front of current in the
Yang-Mills equation (2.2).

We introduce ${\bf E}_{{\rm Q}}^a$ and ${\bf j}_{{\rm Q}}^a$
in the following form
$$
{\bf E}_{{\rm Q}}^a ({\bf x}, t) =
\int \, {\bf E}_{{\bf k}, \omega}^a {\rm e}^{i {\bf k} {\bf x} - i \omega t}
\, {\rm d} {\bf k} {\rm d} \omega , \;
{\bf j}_{{\rm Q}}^a ({\bf x}, t) =
\int \, {\bf j}_{{\bf k}, \omega}^a {\rm e}^{i {\bf k} {\bf x} - i \omega t}
\, {\rm d} {\bf k} {\rm d} \omega .
\eqno{(A5)}
$$
The Fourier-component of a field ${\bf E}_{{\bf k}, \omega}^a =
{\bf k} E_{{\bf k}, \omega}^a/ \vert {\bf k} \vert$ is associated with
${\bf j}_{{\bf k}, \omega}^a$ by Yang- Mills equation
$$
E_{{\bf k}, \omega}^a = \frac{i}
{\omega \varepsilon^l ( \omega, {\bf k} )}
\; \frac{({\bf k} \cdot {\bf j}_{{\bf k}, \omega}^a)}
{\vert {\bf k} \vert} .
\eqno{(A6)}
$$
Substituting $(A5)$ and $(A6)$ into $(A4)$ and taking into account reality of
${\cal W}^l$, we obtain
$$
{\cal W}^l =  \frac{(2 \pi)^3}{2} \int \, i
\left( \frac{1}{\omega^{\prime} \varepsilon^l (\omega^{\prime}, {\bf k})}
- \frac{1}{\omega \varepsilon^{\ast l} (\omega, {\bf k})} \right)
\, \frac{k^i k^j}{{\bf k}^2}
\langle j_{{\bf k}, \omega}^{\ast a i} \,
 j_{{\bf k}, \omega^{\prime}}^{a j} \rangle
{\rm e}^{- i (\omega^{\prime} - \omega) t} {\rm d} \omega
{\rm d} \omega^{\prime} {\rm d}
{\bf k} .
\eqno{(A7)}
$$
Here, by random character of a current phase, we replace the product
$j_{{\bf k}, \omega}^{\ast a i} \,
j_{{\bf k}, \omega^{\prime}}^{a j}$ by its averaged value.

The colour current formed by the constrained motion charge
$(A3)$ is presented as
$$
{\bf j}_{\rm Q}^a({\bf x},t) = g {\bf v} (t) {\rm Q}^a (t) \,
\delta ({\bf x} - {\bf x} (t))
= g {\bf v} (t) {\rm Q}^a (t)
\int \, {\rm e}^{i {\bf k} ( {\bf x} -  {\bf x} (t))}
\frac{{\rm d} {\bf k}}{(2 \pi)^3} = \int \, {\bf j}_{{\bf k}}^a (t) \,
{\rm e}^{i {\bf k} {\bf x}} {\rm d} {\bf k} ,
$$
where we have in the linear approximation
$$
{\bf j}_{\rm k}^a(t) \equiv \frac{g}{(2 \pi)^3} {\bf v} (t) {\rm Q}^a (t)
\, {\rm e}^{- i {\bf k} {\bf x} (t)} \simeq
\eqno{(A8)}
$$
$$
\simeq \frac{g}{(2 \pi)^3} {\bf v}_0 {\rm Q}_0^a
\, {\rm e}^{- i {\bf k} {\bf v}_0 t} +
\frac{g}{(2 \pi)^3} \{ {\rm Q}_0^a [ \Delta {\bf v} (t) - i {\bf v}_0
( {\bf k} \cdot \Delta {\bf x} (t) ) ] +
{\bf v}_0 \Delta {\rm Q}^a (t) \} {\rm e}^{- i{\bf k} {\bf v}_0 t} ,
$$
and $\Delta {\bf x} (t), \Delta {\bf v} (t), \Delta {\rm Q}^a$ are defined
by $(A3)$.

The first term in the r.h.s. of $(A8)$ is connected with linear Landau damping
which is absent in QGP.
Because of this fact this term is omitted.
The term in braces, proportional to
${\rm Q}_0^a$ yields the Abelian contribution to radiation and is connected with
usual spatial oscillation of a colour particle. The term with
$\Delta {\rm Q}^a$ gives the non-Abelian part of radiation and is
induced by precession of a colour vector of a
particle in the field of incident wave. As it is easily to see the interference
of these two contribution vanishes. Further we shall restrict our consideration
to the second part of radiation which induced by non-Abelian part of
a colour current
$$
( {\bf j}_{{\bf k}, \omega}^a)_{non-Abelian} =
\int \, ({\bf j}_{\bf k}^a (t))_{non-Abelian} \,
{\rm e}^{i \omega t} \frac{d t}{2 \pi} =
$$
$$
= \frac{i g^2}{(2 \pi)^3} f^{abc} {\rm Q}_0^c \int \,
\frac{{\bf v}_0 ({\bf v}_0 {\bf A}_{{\bf k}^{\prime}}^b)}
{\omega_{{\bf k}^{\prime}}^l - {\bf v}_0 {\bf k}^{\prime}}
\, \delta ( \omega - \omega_{{\bf k}^{\prime}}^l + {\bf v}_0 ({\bf k}^{\prime} -
{\bf k})) \, {\rm d} {\bf k}^{\prime} .
$$
Substituting the last expression into $(A7)$, taking into account
$$
\langle A_{{\bf k}^{\prime}}^{\ast b}  A_{{\bf k}^{\prime \prime}}^d \rangle =
I_{{\bf k}^{\prime}}^l \delta ( {\bf k}^{\prime} - {\bf k}^{\prime \prime})
\delta^{b d}
$$
and integrating over ${\rm d} \omega ,
{\rm d} \omega^{\prime}$
and ${\rm d} {\bf k}^{\prime \prime}$, we obtain with replacement of argument
${\bf k}^{\prime} \rightarrow {\bf k}_1$
$$
( {\cal W}^l)_{non-Abelian} =
- \frac{g^4N_c \, q^2}{(2 \pi)^3}
\int \frac{{\rm d}{\bf k}{\rm d}{{\bf k}_1}}
{{\bf k}^2 {\bf k}^2_1} \,
{\rm Im} \Big( \frac{1}{\tilde{\omega}^l_{{\bf k}_1}
\varepsilon^l( \tilde{\omega}^l_{{\bf k}_1},{\bf k})} \Big)
\frac{({\bf k}{\bf v})^2 ({{\bf k}_1}{\bf v})^2}
{( \omega^l_{{\bf k}_1} - {\bf v}{{\bf k}_1})^2} \,
I^l_{{\bf k}_1} .
\eqno{(A9)}
$$
Here, $q^2 \equiv {\rm Q}^a_0 {\rm Q}^a_0, \,
\tilde{\omega}^l_{{\bf k}_1} \equiv \omega^l_{{\bf k}_1}
- ({\bf k}_1 - {\bf k}){\bf v}$
and the suffix "0" of velocity is omitted.

For imaginary part of longitudinal permittivity we use approximation
similar to [21]
$$
{\rm Im} \frac{1}{\varepsilon^l(\tilde{\omega}^l_{{\bf k}_1}, {\bf k})}
\simeq - \pi \Big(
\frac{\partial {\rm Re} \, \varepsilon^l(k)}
{\partial \omega} \Big)^{-1}_{\omega = \omega^l_{\bf k}}
\delta( \omega^l_{\bf k} - \tilde{\omega}^l_{{\bf k}_1}).
$$
Going from the spectral density $I^l_{{\bf k}_1}$ to density of longitudinal
oscillations number $N^l_{{\bf k}_1}$ and setting the constant $q^2$
equal to $2/(2 \pi)^3$, we can cast radiaton intensity expression (A9)
into the final form
$$
( {\cal W}^l)_{non-Abelian} =
\int \, {\it w}^c_{\bf v}({\bf k},{\bf k}_1)
N^l_{{\bf k}_1} \omega^l_{\bf k}
\frac{{\rm d}{\bf k}{\rm d}{{\bf k}_1}}
{(2 \pi)^6},
\eqno{(A10)}
$$
where ${\it w}^c_{\bf v}({\bf k},{\bf k}_1)$ is defined by (9.3).
Expression $(A10)$ represents radiation intensity of isolated colour charge
moving in a quark-gluon plasma in the direction to ${\bf v}$ of a field
of a longitudinal wave
with frequency $\omega^l_{\bf k}$ and wave vector ${\bf k}$.

\end{document}